\documentclass[acmsmall]{acmart}

\AtBeginDocument{%
  \providecommand\BibTeX{{%
    \normalfont B\kern-0.5em{\scshape i\kern-0.25em b}\kern-0.8em\TeX}}}

\setcopyright{acmcopyright}
\acmJournal{PACMHCI}
\acmYear{2021} \acmVolume{5} \acmNumber{CSCW1} \acmArticle{80} \acmMonth{4} \acmPrice{15.00}\acmDOI{10.1145/3449154}

\acmConference[CSCW]{PACM HCI}{2021}{}
\usepackage{colortbl}
\usepackage{pdflscape}
\usepackage{graphicx}
\usepackage{capt-of}
\usepackage{multirow}
\usepackage{xcolor}
\usepackage{soul}

\newcommand{\hlc}[2][yellow]{{%
    \colorlet{foo}{#1}%
    \sethlcolor{foo}\hl{#2}}%
}
\newcommand\blfootnote[1]{%
  \begingroup
  \renewcommand\thefootnote{}\footnote{#1}%
  \addtocounter{footnote}{-1}%
  \endgroup
}

\begin{document}

\title{Learning to Ignore: A Case Study of Organization-Wide Bulk Email Effectiveness}

% \author{Leave Authors Anonymous}
% \email{email}
% \affiliation{%
%   \institution{Institute}
%   \city{City}
%   \state{Address}
% }
% \author{Leave Authors Anonymous}
% \email{email}
% \affiliation{%
%   \institution{Institute}
%   \city{City}
%   \state{Address}
% }

% \author{Leave Authors Anonymous}
% \email{email}
% \affiliation{%
%   \institution{Institute}
%   \city{City}
%   \state{Address}
% }

\author{Ruoyan Kong}
\email{kong0135@umn.edu}
\affiliation{%
  \institution{University of Minnesota - Twin Cities}
%   \city{Twin Cities}
  \state{USA}
}

\author{Haiyi Zhu}
\email{haiyiz@cs.cmu.edu}
\affiliation{%
  \institution{Carnegie Mellon University}
%   \city{City}
  \state{USA}
}

\author{Joseph A. Konstan}
\email{konstan@umn.edu}
\affiliation{%
  \institution{University of Minnesota - Twin Cities}
%   \city{City}
  \state{USA}
}


\begin{abstract}
Bulk email is a primary communication channel within organizations, with all-company emails and regular newsletters serving as a mechanism for making employees aware of policies and events. Ineffective communication could result in wasted employee time and a lack of compliance or awareness. Previous studies on organizational emails focused mostly on recipients. However, organizational bulk email system is a multi-stakeholder problem including recipients, communicators, and the organization itself. 

We studied the effectiveness, practice, and assessments of the organizational bulk email system of a large university from multi-stakeholders' perspectives. We conducted a qualitative study with the university's communicators, recipients, and managers. We delved into the organizational bulk email's distributing mechanisms of the communicators, the reading behaviors of recipients, and the perspectives on emails' values of communicators, managers, and recipients.

We found that the organizational bulk email system as a whole was strained, and communicators are caught in the middle of this multi-stakeholder problem. First, though the communicators had an interest in preserving the effectiveness of channels in reaching employees, they had high-level clients whose interests might outweigh judgment about whether a message deserves widespread circulation. Second, though communicators thought they were sending important information, recipients viewed most of the organizational bulk emails as not relevant to them. Third, this disagreement was amplified by the success metric used by communicators. They viewed their bulk emails as successful if they had a high open rate. But recipients often opened and then rapidly discarded emails without reading the details. Last, while the communicators in general understood the challenge, they had a limited set of targeting and feedback tools to support their task.

\end{abstract}



\keywords{organizational communication, email}

\maketitle

\section{Introduction}

\blfootnote{This is a pre-print version of a paper accepted to CSCW 2021 — the 24th ACM Conference on Computer-Supported Cooperative Work and Social Computing.}
Bulk email is email that is sent to a large group of recipients \cite{doi:10.1080/02680513.2018.1556090}. Bulk email has become an important part of organizational communication --- the collective and interactive process of generating and interpreting information within organizations to achieve their purposes  \cite{stohl1995organizational}.
 Organizations (entities comprising multiple people with a particular purpose \cite{handy2007understanding}) often use bulk emails to deliver information to their employees \cite{10.1145/1180875.1180941}.  Example organizational bulk emails include announcements of new staff, summaries of meetings, health and safety issues, and event invitations to relevant groups within organizations, etc. Despite its important role in organizational communication, organizational bulk emails can bring email overload to the organizations' employees. In large organizations, the problem of overwhelming communication is widespread. Recipients are receiving emails irrelevant to them while missing important ones \cite{doi:10.1080/02680513.2018.1556090,Jackson:2003:UEI:859670.859673}.

Previous research on organizational emails mainly studied general emails from recipients' perspectives, such as which emails employees would delay reading \mbox{\cite{Sarrafzadeh2019CharacterizingAP}}, the probability of an email being retained/deleted \mbox{\cite{10.1145/765891.766073}}, \mbox{recipients’} productivity \mbox{\cite{mark2016email}}. However, bulk email communications within organizations is not only a problem involving email recipients but is also an example of a multi-stakeholder problem \cite{andriof2002introduction}. The stakeholders in organizations include:
\begin{itemize}
   \item Communicators: the staff who are in charge of designing and distributing organizational bulk emails.
  \item Communicators' clients: the original producer of organizational bulk emails, who are the communicators' internal clients.
   \item Recipients: the employees who receive organizational bulk emails from the communicators.
\end{itemize}

 Not only are there different goals for communicators, their clients, and recipients, but also there is a key fourth stakeholder --- the organization itself --- that has its priorities not always recognized by communicators and their clients, or recipients.  Communicators and their clients naturally focus on their own needs --- getting the word out, establishing evidence of notice or compliance, or preserving a record of communication. But recipients, faced with more communication than they can handle, have to scan, filter, or simply ignore organizational bulk emails. In turn, organizational goals around compliance, informed employees, and employee productivity may suffer.

% This multi-stakeholder system relies on communicators' efforts to design and distribute organizational bulk emails. For example, in the University of XXX, an organization with several campuses, over 25,000 employees and 50,000 students, there are 44 employees who are in charge of university-wide communication. How to design and distribute organizational bulk emails effectively across one organization with this organization's communicator resources remains unsolved. 

 Thus we would like to understand multi-stakeholders' perspectives within the organizational bulk email system. We conducted a case study, a predominant approach in organizational studies \mbox{\cite{swanborn2010case, buchanan2012case}} and organizational communication studies \mbox{\cite{Sarrafzadeh2019CharacterizingAP, 10.1145/765891.766073, mark2016email, SCHULER1979268, stohl1987messages}}. A case study could help us 1) learn domain experts' (communicators) information management techniques and how the bulk email system is used within the organization context; and 2) understand the links between different stakeholders within the same organization \mbox{\cite{lazar2017research}}. We selected a representative organization with hierarchical structures, and centralized and decentralized communication offices. We hope our case study will 1) inform the design of the organizational bulk email system \mbox{\cite{yin2003case}}; 2) provide interview protocols for future studies on the system; and 3) be generalizable to similar organizations.
% Research Methods in Human-Computer Interaction 2nd Edition 4.5 star rating 2 Reviews Authors: Jonathan Lazar Jinjuan Feng Harry Hochheiser Paperback ISBN: 9780128053904 Chapter 7 -- Case Studies 2nd

Specifically, we studied communicators' practice of designing and distributing organizational bulk emails, and communicators/recipients/managers' experience and assessments. We combined self-reported data, logged inbox data, and inbox-review data in an artifact walkthrough \cite{10.1145/227614.227615}.

 We found that the organizational bulk email system as a whole was strained, and showed its strain in the intermediary role of the communicator. First, communicators had high-level clients who wanted their messages sent broadly. Second, communicators did not have visibility into the effectiveness of bulk emails in eliciting behaviors or transmitting information; they only measured bulk emails by ``open rates''. Communicators felt they were sending important information and got high ``open rates''. Third, recipients viewed most bulk emails as not relevant to them; they often opened and then rapidly discarded emails without reading the details. Last, while the communicators in general had an interest in preserving the effectiveness of channels in reaching employees, they had a limited set of designing, targeting, and feedback tools to support their task.

The rest of this paper includes related work (2), gaps and research questions (3), study site and its bulk email system's structure (4), artifact
walkthrough with stakeholders (5, 6), and discussion (7).

\section{Related Work}
\subsection{Organizational Communications}
% Communication within organizations has been studied for more than a century \cite{van2012electronic}. 
Communication within organizations has been studied for more than a century \cite{van2012electronic}. Communication has been called \textit{``the life blood of organization''} \cite{goldhaber1990organizational}, \textit{``the glue that binds it all together''} \cite{katz2008communication}. Myers and Myers \cite{myers1982managing} defined organizational communications as “the central binding force that permits coordination among people and thus allows for organized behavior''.

 Organizational communication is a multi-stakeholder problem. Organizational communication's effectiveness is a problem not only about information recipients, but also information providers (like communicators and their clients). Randall \mbox{\cite{SCHULER1979268}} did a case study in a large manufacturing firm, finding that if information producers failed to provide clear information, it was unlikely that the information receivers would continue to seek out information. Stohl and Redding \mbox{\cite{stohl1987messages}} studied a Chinese steel construction corporation, finding that workers' cognitive failures were significantly related to the managers' communication styles.

Organizational communication's effectiveness is also related to the organization itself. Greenbaum \cite{greenbaum1974audit} defined the purpose of organizational communications as the achievement of organizational goals, accomplished through the appropriate employment of communication networks, policies, and activities. Greenbaum argued that communication effectiveness had to be measured by looking at the overall communication system besides the activities of recipients. 

  Ineffective organizational communication is often linked to the behaviors of multiple stakeholders, not just recipients. For example, high-level senders did not provide enough information about what was going on at the organization, gatekeepers (e.g. communicators) lost/delayed delivery of some vital information in the distribution process, or recipients' feedback was neglected \mbox{\cite{downs2012assessing, odine2015communication}}.

 However, as we discussed below, as a part of the organizational communication studies, the previous studies on organizational emails focused mostly on recipients and less on information providers and the organization itself.

% The widely-application of emails brought changes to organizational communication. Email enabled sender and recipients to control the timing of their portion of the communication, speed up the exchange of information and leaded to the exchange of new information \cite{sproull1986reducing}, expedited communication frequency \cite{feldman1987electronic}, created what Sproull and Kiesler called a “networked organization” \cite{sproull1991computers} in which people can be available even when they are physically absent. However, the widely-application of emails in workplace also brought email overload.
% To our surprise, there are few tools specifically designed to support employee email communication within organizations (both we and our IT department conducted searches), though there is a widely used set of discussion/archive and instant messaging tools such as Basecamp and Slack.  By contrast, there is a wealth of tools designed to support sending email messages to external clients (tools on the marketing/customer relationship management spectrum such as MailChimp, ConstantContact, and Salesforce).  In our preliminary work we have found that numerous institutions (including our own) are starting to adopt such tools for internal use, presenting opportunities for query-based recipient selection and other forms of within-message personalization (traditional mail-merge from tables, but also selection of one of several components for newsletters).  These tools at best support only sender utility, and do nothing to support recipient or organizational utility.
\subsection{Organizational Email}
 Organizational internal emails (we refer to them as organization emails) are emails whose senders and recipients are within the same organization. Organizational emails create what Sproull and Kiesler called a “networked organization” \cite{sproull1991computers}, in which people can be available even when they are physically absent. At the same time, they bring email overload --- email users’ perceptions that their own use of
email has gotten out of control because they receive more emails than they can handle effectively \cite{10.1145/1180875.1180941, fisher2006revisiting,grevet2014overload, whittaker1996email}. 

 Most studies of organizational emails focused on recipients and recipient-perceived effectiveness. First, previous studies around estimating organizational email effectiveness used data which only contained one kind of stakeholder --- recipient's log data/inbox data/behavior data \cite{Sarrafzadeh2019CharacterizingAP, paczkowski2016checking}, like the Avocado dataset \cite{yang2017characterizing}, Enron Corpora \cite{bekkerman2004automatic, bermejo2011improving, klimt2004enron,6970931}, and Outlook Dataset \cite{alrashed2019evaluating}.

Second, previous studies around email overload focused on how email burdened its recipients, instead of the organization \cite{bellotti2005quality, mishra2014driving, malone1987information}. Dabbish and Kraut \cite{10.1145/1180875.1180941} conducted a study that quantified the email burden on white-collar workers. Merten and Gloor \cite{merten2010too} found that employee job satisfaction went down as internal email volume increased in a case study in a 50 people company. Huang and Lin \cite{huang2009factors} surveyed 3 universities and found that knowledge workers were “ruled by email”. 

Third, previous studies mostly assumed that information providers (communicators and their clients) should change their email actions according to recipients' needs to reduce email overload. Jackson et al. \cite{Jackson:2003:UEI:859670.859673} studied 16 employees at the Danwood Group and proposed that email frequency should be controlled. Lu et al. \cite{lu2012epic} described EPIC, an email prioritization tool which combined global priority with individual priorities.  Reeves et al. \cite{reeves2008marketplace} experimented with attaching virtual currency to emails to signal importance in a large company.

\subsection{Bulk Email}
Bulk email is email sent to a large group of subscribers for content delivery \cite{doi:10.1080/02680513.2018.1556090}. Bulk email is a primary communication infrastructure within organizations, businesses and  industries. 

% Bulk email is used by retailers to send advertisements to customers for marketing needs \cite{Lee:2018:ITM:3170427.3188684} outside of organizations; it functions as task assignment and newsletter distribution inside of organizations \cite{Jackson:2003:UEI:859670.859673}.

Most studies of bulk email are about external bulk emails --- the emails sent to recipients outside of organizations, like marketing emails or  advertisements. These include: 1) supporting information providers design external bulk emails: Carter et al. \cite{carter2011using} found that bulk email was more effective when it was entertaining. Trespalacios and Perkins \cite{trespalacios2016effects} examined the effects of bulk email designs and found that neither the degree of personalization nor the length of the invitation email impacted survey response; 2) supporting recipients filter external bulk emails: Gray and Haahr presented an architecture for personalized, collaborative spam filtering by content-based approaches \cite{gray2004personalised}, and reached over $90\%$ accuracy in a 2-week case study. Al-Jarrah et al. proposed header-based approaches \cite{al2012identifying}, reaching a ROC Area of 98.5\% in CEAS2008 dataset.

\section {Gaps and Research Questions}
We identified 2 major gaps from the previous work:
\begin{itemize}
    \item  The studies on organizational emails focused on recipients: though studies on organizational communication pointed out that when measuring organizational \mbox{communication’s} effectiveness we should consider not only recipients, but also information providers (communicators and their clients) and the organization itself, previous studies on organizational emails did not learn whether recipients and information providers within an organization had different preferences about emails, and how these mismatches affected the organization.
    \item  The studies on organizational bulk emails focused on external bulk emails: few works on bulk emails considered the case when the bulk email sender and  recipients were in the same organization, and when the ultimate goal was maximizing this organization's interests instead of the information providers' or the recipients' interests. Within organizational context, whether a recipient (employee) should get an organizational bulk email may not be determined by whether the recipient likes the email. Sometimes employees may have responsibility for knowing about this email though they might not have interests in it. We see a need to understand bulk emails from an organizational context. 
\end{itemize}

Therefore we identified a need to undertake a systematic investigation of bulk email communications from a whole organization's view. Notice that we \hlc[white]{referred} to organizational bulk email as \textbf{bulk email} and organizational bulk email system as \textbf{bulk email system} in the rest of this paper. We \hlc[white]{posed} the following research questions:

\noindent\textbf{Q1:}  What are communicators' current practices for designing and distributing bulk emails? 

\noindent\textbf{Q2:}  What are the experience with and assessments of bulk emails of different stakeholders?

We conducted an in-depth case study of a representative organization. We interviewed both communicators and recipients within an organization using an artifact walkthrough approach. We delved into email cases and got their assessments of email values. \hlc[white]{We discussed} communicators, recipients, communicators' clients, and recipients' managers in this paper.

\section{Study Site and Its Bulk Email System's Structure}
\subsection{Organization Environment}
The organization \hlc[white]{we studied} in this paper is \hlc[white]{the University of Minnesota} --- a large university with more than 25,000 employees and several campuses. It is structured much like many large universities, with \textit{``a mixture of collegiate, managerial, senatorial, centralized and decentralized governing styles''} \cite{dill1995emerging}: 
\begin{enumerate}
    \item  There \hlc[white]{were} central units led by university leaders (Office of President, University Relations, Information Technology, etc). They were responsible for the core business functions.
    \item  There \hlc[white]{were} decentralized units, like colleges and campuses led by academic leaders (deans and chancellors) and organized into departments. Most of these offices performed a piece of centralized functionality locally, with own human resource, information technology, communicators. These offices also worked with the centralized offices for other functions.
    \item  Each of the units had a communication office, with communicators (communication director/staff), organized according to the location of the units. There were central communication offices in central units and decentralized communication offices in decentralized units.
\end{enumerate}

 The decentralized and centralized structures might cause the segmentation of values and responsibilities of units, as Kuh and Whitt found \mbox{\cite{kuh2000culture}}: \textit{``Universities are not monolithic entities. Subgroups have their own artifacts and values, which may differ from the host''}. For example, faculty felt a responsibility not only to their university but also to their discipline/program \mbox{\cite{masland1985organizational}}. These
conflicting responsibilities led to the decreased potential for coordination \mbox{\cite{shriberg2002sustainability}}.

\subsection{Bulk Email System's Structure}
\begin{figure}[!htbp]
\centering
  \includegraphics[width=0.9\columnwidth]{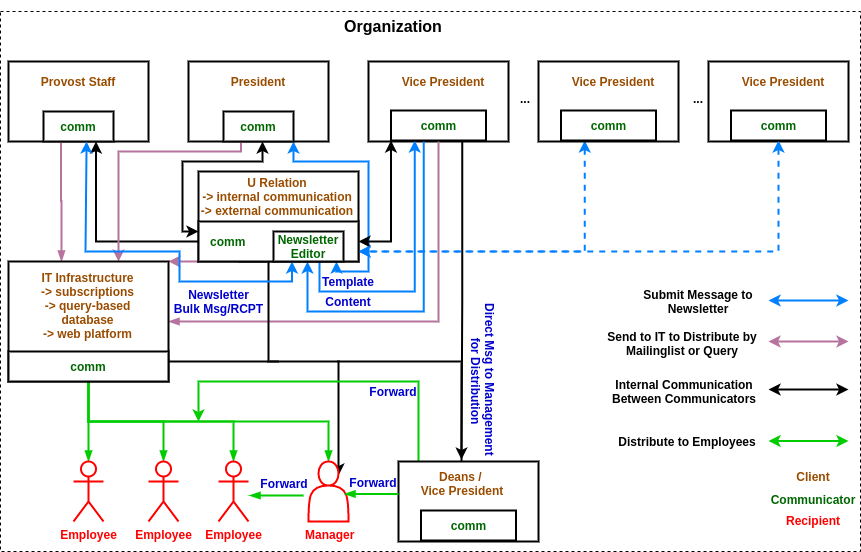}
  \caption{Structure of the Bulk Email System of the University; ``comm'' stands for communicators. }~\label{fig:comm}
\end{figure}

We met with a group of 9 communicators within this university. In the meeting, we found agreement that the current bulk email system might not be what the university wants it to be, since the communicators felt that bulk emails were ignored too often. They encouraged us to move forward. Fig \ref{fig:comm} is the structure of the bulk email system of the university:
\begin{enumerate}
   \item   They used a Gmail system based on G Suite for Education \mbox{\cite{celik2018importance}} with a Salesforce platform, and a domain-based email authentication system to identify legitimate bulk emails.
    \item   The central communication offices conducted university-wide email communications. The decentralized communication offices worked with the central communication offices and conducted local bulk email communications. 
    \item   Only the trained communicators in each unit can send and manage the targeting of bulk emails. Senior communicators were also responsible for drafting emails for university leaders.
    \item   Communication offices submitted bulk email content to the newsletter editors in  offices like University Relations. Employees would be subscribed to these newsletters by default.
    \item   The access to sending bulk emails to large campus-wide or system-wide groups through existing mailing lists, or pulling out lists from PeopleSoft's database (a Human Resource System) was controlled by the central information technology services.
    
\end{enumerate}

% Each working unit had a communication office with at least one communicator, some including a communicator manager and several communicator staff. They were in charge of this unit's communication, including designing and distributing OBEs. There were centralized communication offices (the communication offices in Provost, President, Vice President, University Relations' Unit) and decentralized communication offices (the communication offices in colleges and departments). The centralized communication offices conducted university-wide communications. Specifically, the University Relations Unit was in charge of the internal communications between centralized communication offices. The centralized communication offices sometimes sent OBEs directly to employees by using existing mailing lists, or pulling out lists from Salesforce's database (we will discuss the methods of targeting later); other times they used a distribution mechanism where the OBEs were sent to the decentralized communication offices, and asked them to distribute the OBEs to their employees selectively.
% The communicators between different working units also worked together. For example, besides the normal OBEs, there are newsletters, which were sent to their subscribers daily/weekly/monthly. Employees would be subscribed to these newsletters by default without being informed. Some of these mailing lists could be unsubscribed and some could not be opt-out. Besides sending OBEs themselves, communicators also send content to newsletter editors. The content would be put into the templates designated by newsletter editors. Then newsletter editors checked and modified, put them into newsletters and sent out.

Each bulk email contained one or more messages. We \hlc[white]{defined} a \textbf{message} as a single cohesive amount of information; for the bulk email with several messages, a message is usually a subsection of it. Where participants \hlc[white]{used} ``message'' differently from our definitions, we \hlc[white]{clarified} in the quotes with bracketed notes \textit{``I scan for urgent messages [emails].'' (R5)}

Among the organizational emails of this university (the emails sent from the university's email address \hlc[white]{@umn.edu}), according to the channels and the number of messages of the emails, we \hlc[white]{identified} three categories: 
\begin{itemize}
    \item \textit{Non-Bulk Emails:} The organizational emails not sent by bulk email channels, like individual emails to other employees or to small groups without a mailing list.
    \item \textit{Mass Emails:} The bulk emails with a single message. For example, an email sent to all employees to announce a free financial workshop from the human resources office.
    \item \textit{Newsletter Emails:} A special kind of bulk email that \hlc[white]{had} a collection of messages and \hlc[white]{was} sent to the recipients periodically. For example, a weekly news digest sent to all employees, featuring administrative, legislative, budgetary, event, and other information.
\end{itemize}

% How well did this bulk email system within this university work and what were the practice, experiences, and assessments of the different stakeholders within it? We conducted a survey study (section 5) and an artifact walkthrough study (section 6) within this university to answer these questions. 

\section{Artifact Walkthroughs with Stakeholders --- Methods}
  What are the practices, experiences, and assessments of different stakeholders of the bulk email system within this university? We conducted artifact walkthroughs with 17 stakeholders --- 6 communicators, 9 recipients, and 2 managers within the university. Artifact walkthrough is used to walk the interviewer through the process of completing a task. It helps the interviewer infer possible design intent by
asking probing questions about why one thing was done rather
than another \cite{beyer1999contextual}. In our study, for example, recipients used certain queries in their inbox to retrieve specific types of emails; then answered questions about their practice (trashed/opened/scanned/read in detail) and assessments (sometimes they were asked to reread the email) on each email case retrieved.

\subsection{Recruitment and Participants}
We interviewed 17 participants in the same university by the following process:
\begin{enumerate}
    \item   We worked with the same group of communicators in section 4.2 to generate a list of potential participants by a stratified sampling approach,  which included 8 communicators and 20 recipients, based on length of work experience, nature of the office, and job responsibility.
    \item   We started inviting them for a 30 to 60 minutes interview in our lab or their own offices.
    \item   We stopped inviting after we interviewed 6 communicators and 9 recipients because we had stopped finding new observations from adding more communicators/recipients.
    \item   To avoid recipients' answers being influenced by their awareness of manager involvement, we waited till after the interview to ask the recipient if they would allow us to invite their direct manager to discuss the non-personal emails we collected from the recipient. Two recipients agreed and we invited their managers for an extra 30-minute interview.
\end{enumerate}

% To study to what extent OBEs worked to support this university's daily work,

This study was reviewed and approved by the IRB of the University of Minnesota. We had 17  participants ( Table \ref{tab:demographic}): 1) there were 6 communicators --- 2 communication directors who had over 10 years of experience and 4 communication staff, from 5 different central offices and 1 college office; their responsibilities included editing newsletter, writing drafts for university leaders, leading technical support, etc; 2) there were 9 recipients --- 5 staff (1 from a central office, 2 from college offices, 2 from departmental offices) and 4 faculty; 5 recipients had been at the university for over 10 years; 3) R5's manager (a college office's director) and R6's manager (a program director) were also invited.

\subsection{Interview Protocol}
\noindent\textbf{A. Communicators.} The artifact-walkthroughs with communicators included 3 parts: 
\begin{itemize}
\item General practice questions on their duties, goals, and mechanisms of distributing bulk emails.
\item Email case questions on how an email was designed, sent and measured. We asked communicators to select important/unimportant cases from their points of view.
\item General assessment questions, see Table ~\ref{tab:gatekeeper_protocol} in the appendix.
\end{itemize}

\noindent\textbf{B. Recipients.} The artifact-walkthroughs with recipients included 3 steps: 
\begin{itemize}
    \item {Collect inbox logged data: They were asked to copy and paste 10 email queries to retrieve subsets of organizational emails, non-mass emails, mass emails, and newsletters they received in the past week. We recorded the number of emails left unread/opened of each category (see the actions' definitions in Table \ref{tab:action_definition} --- logged left unread/opened).
    
      Notice that we retrieved emails in all categories, including those archived and in trash/spam, by command ``in:anywhere'' (see Table \mbox{\ref{tab:command}}), except those permanently deleted from the trash. By default, only emails that have been in the trash for more than 30 days would be automatically deleted by Gmail. We did not observe the case that the trash was manually cleaned by the recipients. We assumed that we could retrieve all emails received in the past 7 days.
    
      These queries were pretested in one of the author's inbox to ensure they retrieve the right subset of emails. We recorded the size of each subset, and how many were opened.}
    \item Collect self-report data: For the mass/newsletters we retrieved above, we investigated up to 12 emails for each category. For each email, we recorded the recipient's actions on: 1) whether the email was left unread/opened/scanned/read in detail/trashed (see the actions' definitions in Table \ref{tab:action_definition} --- self-report left unread/opened/ scanned/read in detail/trashed); 2) whether asked for actions/took actions; 3) importance/urgency/relevance on scale 1 to 5; 4) reasons for the answers/actions, and 5) whether they would change their mind about answers/actions if they reread it now (see appendix B). We investigated 163 cases in total.
    \item General questions: We asked them some general questions like how frequently they checked their email accounts, how many emails accounts they had, did they feel that it was too many to read all of them, and how often did they not read an email. 
\end{itemize}

% We were able to do this because of the centralization of bulk email systems; the emails sent to a mailing list from the university would come from addresses ended with XXX.edu, and have ``list'', ``lists'' in the ``to'' field, or ``local'' in the ``list'' field, or be classified into forums or promotions categories in the university's Gmail system.

\noindent\textbf{C. Managers.} The answers from the employee-manager pair were confidential from each other. We showed the managers the emails we collected from their employees, and asked them to give their preferred actions that their employees ought to have done with those emails --- trashed/left unread/opened/scanned/read in detail, and estimations on importance/urgency from 1 to 5. We investigated 31 emails in total with the two managers. \hlc[white]{Notice that the ``manager'' here means the direct supervisor of the recipients, not a reflection of a particular position in the organization. And 
the recipients listed as "director" in Table \mbox{\ref{tab:demographic}} are heads of their offices or programs but are not listed as "manager" because we interviewed them about their own received messages, not about the messages of one of their employees.} 

\hlc[white]{To summarize, for participants invited as communicators/recipients/managers, we discussed the bulk emails they sent/received/their employees received correspondingly.}

\begin{table}[!htbp]
\centering
\resizebox{\textwidth}{!}{%
\begin{tabular}{|c|c|} 
\hline
 \textbf{The Actions of Recipients}  & \textbf{Definition}                                                                                                                                                                                                 \\ 
\hline
Logged Left Unread                   & The email was labeled unread by Gmail.                                                                                                                                                                              \\ 
\hline
Self-Report Left Unread              & The recipient claimed that they did not open the email.                                                                                                                                                             \\ 
\hline
Logged Opened                        & The email was labeled read by Gmail.                                                                                                                                                                                \\ 
\hline
Self-Report Opened                   & The recipient claimed that they opened the email.                                                                                                                                                                   \\ 
\hline
Self-Report Scanned                  & The recipient claimed that they read the email quickly to get its general idea only.                                                                                                                                \\ 
\hline
Self-Report Read in Detail           & The recipient claimed that they read the whole email thoroughly and carefully.                                                                                                                                      \\ 
\hline
Self-Report Trashed                  & \begin{tabular}[c]{@{}c@{}}The recipient claimed that they archived the email or moved it to the trash folder; \\untrashed email was left in the inbox or moved to a non-trash folder.\\ \end{tabular}  \\ 
\hline
Self-Report Opened and Trashed       & The recipient claimed that they opened the email first then trashed it.                                                                                                                                             \\
\hline
\end{tabular}}
\caption{The definitions of recipients' actions with emails. Logged data was collected by using queries to retrieve subsets. Self-report data was collected by asking recipients their actions directly.}\vspace{-0.3in}
\label{tab:action_definition}
\end{table}

\begin{table}[!htbp]
\centering
\resizebox{\textwidth}{!}{%
\begin{tabular}{|c|c|c|c|c|c|c|c|c|c|} 
\hline
\textbf{\#} & \begin{tabular}[c]{@{}c@{}}\textbf{Stakeholder}\\\textbf{Type}\end{tabular} & \begin{tabular}[c]{@{}c@{}}\textbf{Years at }\\\textbf{University}\end{tabular} & \textbf{Position} & \begin{tabular}[c]{@{}c@{}}\textbf{Level of}\\\textbf{Office}\end{tabular} & \textbf{\#} & \begin{tabular}[c]{@{}c@{}}\textbf{Stakeholder}\\\textbf{Type}\end{tabular} & \begin{tabular}[c]{@{}c@{}}\textbf{Years at}\\\textbf{University}\end{tabular} & \textbf{Position} & \begin{tabular}[c]{@{}c@{}}\textbf{Level of}\\\textbf{Office}\end{tabular}  \\ 
\hline
C1          & Communicator                                                                & 6 - 10                                                                          & Staff             & Central                                                                    & R1          & Recipient                                                                   & 1 - 5                                                                          & Staff             & Central                                                                     \\ 
\hline
C2          & Communicator                                                                & 11- 20                                                                          & Director          & Central                                                                    & R2          & Recipient                                                                   & 1 - 5                                                                          & Staff             & Departmental                                                                \\ 
\hline
C3          & Communicator                                                                & 1 - 5                                                                           & Staff             & College                                                                    & R3          & Recipient                                                                   & 11 - 20                                                                        & Director          & Departmental                                                                \\ 
\hline
C4          & Communicator                                                                & 11 - 20                                                                         & Director          & Central                                                                    & R4          & Recipient                                                                   &  20                                                                            & Staff             & College                                                                     \\ 
\hline
C5          & Communicator                                                                & 1 - 5                                                                           & Staff             & Central                                                                    & R5          & Recipient                                                                   & 1 - 5                                                                          & Staff             & College                                                                     \\ 
\hline
C6          & Communicator                                                                & 6 - 10                                                                          & Staff             & Central                                                                    & R6          & Recipient                                                                   &  20                                                                            & Professor         & \textbackslash{}                                                            \\ 
\hline
M1          & R5's Manager                                                                & 6 - 10                                                                          & Director          & College                                                                    & R7          & Recipient                                                                   & 11 - 20                                                                        & Professor, Director         & \textbackslash{}                                                            \\ 
\hline
M2          & R6's Manager                                                                &  20                                                                             & Professor         & \textbackslash{}                                                           & R8          & Recipient                                                                   & 11 - 20                                                                        & Professor         & \textbackslash{}                                                            \\ 
\hline
\multicolumn{5}{|c|}{}                                                                                                                                                                                                                                                       & R9          & Recipient                                                                   & 1 - 5                                                                          & Professor, Director         & \textbackslash{}                                                            \\
\hline
\end{tabular}}
\caption{Demographics of participants. C1 --- C6, R1 --- R5, M1 are staff and the rest are faculty. Central-level offices are in charge of university-wide affairs, like university services, information technology; college-level or departmental offices are located within a college or department, like the dean's office.}
\label{tab:demographic}\vspace{-0.3in}
\end{table}

\subsection{Data Analysis}
To compare the experience and assessments of the effectiveness of the bulk email system from different stakeholders, we took a grounded theory approach \cite{charmaz2014constructing}, specifically with the following iterative procedure:
\begin{enumerate}
    \item Identifying the themes from the communicators' transcripts.
    \item Searching for relevant text and themes in the recipients' transcripts.
    \item Comparing the actions taken by the recipient and preferred by the manager (if applicable).
    % \item Counting the emails in each subset, to prove the prevalence of the evidence found.
    \item Inviting new participants and repeating the steps above until we reached data saturation. We stopped when we had interviewed 17 stakeholders because we found strong repetition in the themes identified.
\end{enumerate}
 
% The interview transcripts of the communicators were iteratively analyzed to identify emergent themes using a grounded theory approach \cite{charmaz2014constructing}. Initial affinity clusters of data were discussed by three researchers. Successive iterations were completed by one researcher to finalize the central concepts. 

% We also counted the emails in each subsets, to prove the representativeness and prevalence of the evidence in the qualitative data. We presented our results below.

\section{Artifact Walkthroughs with Stakeholders --- Results}
\subsection{Number of Emails Received and Actions Taken}
Table~\ref{tab:email_number} reported the statistics of the 9 recipients on emails received, opened, read, and trashed by category. The recipients, on average, received 376 emails in the past week: 153 of them were organizational emails and 30 of them were bulk emails (25 mass emails and 5 newsletters). The number of messages in one of the 55 newsletters investigated in the interview could be as many as 35, with an average of 8.5. Participants (recipients) reported that they, in general, received too many bulk emails. R6 said \textit{``Sometimes I felt overwhelmed.''}. R4 and R8 said that a large number of bulk emails became a burden to them.

Faculty received 175 organizational emails a week on average (148 non-mass emails, 23 mass emails, 4 newsletters), and for staff, this average number was 136 (103 non-mass emails, 27 mass emails, 6 newsletters). The average logged open rate of mass/newsletters of faculty recipients (>90\%) was higher than their self-report open rate (<70\%). Sometimes they clicked an email's title and removed its unread tag, but they did not think that they ``opened'' the email.

Though many bulk emails got opened, few of them were read in detail. In fact, $58\%$ of mass emails were reported being opened by staff and only $28\%$ of them were read in detail; $67\%$ mass emails were reported being opened by faculty and only $13\%$ of them were read in detail.

Many bulk emails were trashed, $52\%$ mass emails and $22\%$ newsletters were reported being trashed by faculty, for staff the 2 numbers were $27\%$ and $15\%$.

\begin{table*}[!htbp]
\centering
\resizebox{\textwidth}{!}{%
\begin{tabular}{|l|c|c|c|c|c|c|} 
\hline
\multicolumn{1}{|c|}{\textbf{Faculty}} & \textbf{Email Type}                         & \textbf{Overall} & \textbf{Organizational} & \begin{tabular}[c]{@{}c@{}}\textbf{Organizational}\\\textbf{Non-Bulk}\end{tabular}  & \begin{tabular}[c]{@{}c@{}}\textbf{Organizational}\\\textbf{Mass}\end{tabular}  & \begin{tabular}[c]{@{}c@{}}\textbf{Organizational}\\\textbf{~Newsletter}\end{tabular}  \\ 
\hline
\multicolumn{1}{|c|}{no.= 4}           & \textbf{\# Received Per Week}               & 560              & 175                     & 148                                                                                 & 23                                                                              & 4                                                                                \\ 
\hline
                                       & \textbf{\% Logged Open Rate}                       & 55\%             & 96\%                    & 97\%                                                                                & 92\%                                                                            & 94\%                                                                             \\ 
\hline
                                       & \textbf{\% Self-report Open Rate}           & \textbackslash{} & \textbackslash{}        & \textbackslash{}                                                                    & 67\%                                                                            & 59\%                                                                             \\ 
\hline
                                       & \textbf{\% Self-report Read in Detail Rate} & \textbackslash{} & \textbackslash{}        & \textbackslash{}                                                                    & 13\%                                                                            & 9\%                                                                              \\ 
\hline
                                       & \textbf{\% Self-report Trash/Archive Rate}  & \textbackslash{} & \textbackslash{}        & \textbackslash{}                                                                    & 52\%                                                                            & 22\%                                                                             \\ 
\hline
\multicolumn{1}{|c|}{\textbf{Staff}}   & \textbf{Email Type}                         & \textbf{Overall} & \textbf{Organizational} & \begin{tabular}[c]{@{}c@{}}\textbf{Organizational}\\\textbf{~Non-Bulk}\end{tabular} & \begin{tabular}[c]{@{}c@{}}\textbf{Organizational}\\\textbf{~Mass}\end{tabular} & \begin{tabular}[c]{@{}c@{}}\textbf{Organizational}\\\textbf{~Newsletter}\end{tabular}  \\ 
\hline
\multicolumn{1}{|c|}{no. = 5}          & \textbf{\# Received Per Week}               & 229              & 136                     & 103                                                                                 & 27                                                                              & 6                                                                                \\ 
\hline
                                       & \textbf{\% Logged Open Rate}                       & 54\%             & 73\%                    & 78\%                                                                                & 52\%                                                                            & 50\%                                                                             \\ 
\hline
                                       & \textbf{\% Self-report Open Rate}           & \textbackslash{} & \textbackslash{}        & \textbackslash{}                                                                    & 58\%                                                                            & 58\%                                                                             \\ 
\hline
                                       & \textbf{\% Self-report Read in Detail Rate} & \textbackslash{} & \textbackslash{}        & \textbackslash{}                                                                    & 28\%                                                                            & 30\%                                                                             \\ 
\hline
                                       & \textbf{\% Self-report Trash/Archive Rate}  & \textbackslash{} & \textbackslash{}        & \textbackslash{}                                                                    & 27\%                                                                            & 15\%                                                                             \\
\hline
\end{tabular}}
\caption{Average per-capita email volume by category; \% logged open rate is calculated by the inbox logged data; \% self-report open/read-in-detail/trash/archive rate is calculated by the self-report data, based on all the bulk emails investigated in the corresponding category, for example, self-report read in detail rate of mass email = \# mass email read in detail/\# mass email investigated.}
\label{tab:email_number}\vspace{-0.2in}
\end{table*}

\subsection{Nature of Bulk Email System as an Organizational System}
Bulk email system was part of the university's administrative system, with communicators trying to achieve the communication goals of this university. They had clients, performance metrics, and tools.\\

\subsubsection{ Communicators as a gateway for organizational leaders to reach employees.}
~\\
Communicators had clients from their own offices or other offices who were not professionals in communication and wanted to promote events, notify of changes of policies, etc. Many clients were the communicators' leaders:
% \begin{quote}
% \textit{
% Sometimes for university services, for the president's office, provost's office, even central administration...Different offices. I'm in university relations. (C5)}
\begin{quote}
\textit{We work with our leadership whoever sends the message, so for example, we send a few messages for the associate vice president for research, who's also the head of responsive administration. (C1)}
\end{quote}

Naturally, clients wanted to send more content and reach more people. A communicator talked about their clients:

\begin{quote}
    \textit{
We're working with other departments coming up with certain final messages that may be close to the amount of information that people can digest. But some of them think that people have a more of an appetite for reading. (C1)
}
\end{quote}

Thus communicators usually needed to narrow mailing lists and shorten the length of emails. One mechanism they used was letting clients fill in  templates:

\begin{quote} 
 \textit{   The U Relations developed a partnerships with emergency management. When there is a safety alert, there are certain blanks in the template they fill in. That's why you can see the safety messages [emails] are often similar. Because we give them the authority to send them, so we want to limit the scope of the content they can choose to put in there. (C4)
}\end{quote}

Sometimes communicators could not modify the emails. Some bulk emails were required by law:

\begin{quote}
% \textit{We work with our leadership whoever send the message, so for example, we send a few messages from the associate vice president for research. (C1)}

\textit{
     We have things called safety advisories. Maybe somebody is grabbed from the Ford Hall last night and we just found out about it. That person is gone but we are required by federal law to tell you about that. (C4)}\\
\end{quote}

\subsubsection{Communicators are anchored to the ``open rate'' metric}
~\\
Optimizing a system depends on using appropriate metrics. The metrics used by communicators to measure bulk email's effectiveness did not match recipients' assessments. We asked communicators about the metrics they used in the bulk email system, and measured recipients' assessments for each bulk email from various metrics like whether trashed/opened/scanned/read in detail, and ratings on urgency/importance/relevance. For communicators, the priority order of the metrics used to report a bulk email's performance was: open rate > click rate > replies and others. They only needed to report a good open rate to their clients. $50\%$ was a good open rate to communicators:

\begin{quote}    
\textit{For all campus messages [emails], open rates are usually at least 50\%, which is good for us, so pretty high.  We look at click rate but not closely. Most times we look at open rates. (C5)
}\end{quote}

% These open rates were proved from the recipients' side. The probability of being opened was around 60\% for a bulk email, as the faculty/staff self-reported in Table \ref{tab:email_number}. 

\begin{figure}[!htbp]
\centering
  \includegraphics[width=1\columnwidth]{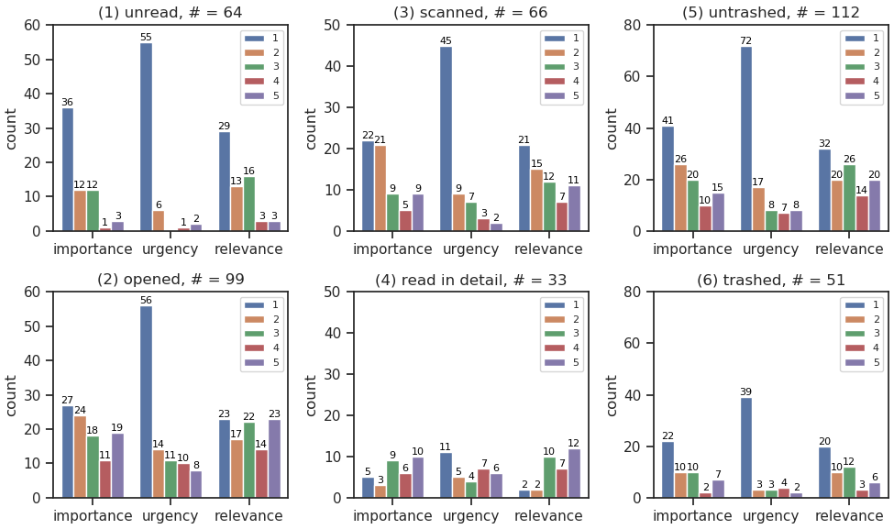}
  \caption{Counts of ratings on importance/urgency/relevance of the investigated bulk emails which were reported being left unread/opened/scanned/read in detail/left untrashed/trashed by the recipients.}~\label{fig:rating_2}
\end{figure}

However, ``open rate'' might be a flawed metric for measuring recipients' feedback, because:
\begin{enumerate}
    \item ``Being opened'' did not mean ``being read''. As we discussed above, the logged open rate, which was used by communicators, might be higher than the self-report open rate --- sometimes recipients clicked an email's title and removed its unread tag, but never read any sentences in the email.  Some faculty recipients, like R7, opened 17 of the 22 bulk emails we investigated, but only read 2 of them in detail, see Table \ref{tab:user_rate}.
    
    Recipients were not opening bulk emails because they considered them relevant but because they were unsure of their relevance and wanted to verify:
    \begin{quote} \textit{
I read the first line then deleted it because I've seen similar information in other news. (R7)
}\end{quote}

    \item Most of ``opened'' bulk emails were only scanned and got low ratings in importance/urgency/ relevance. As shown in Figure \mbox{\ref{fig:rating_2}} (3), 43/54/36 bulk emails \hlc[white]{got} 1 or 2 on importance/urgency/ relevance ratings over 66 scanned emails.
    \item Some bulk emails were opened and then trashed. As shown in Table \mbox{\ref{tab:user_rate}}, R1, R3, R5, R6, R7, R9 all trashed some bulk emails after they opened them. 
    
    Trashing was more like a personal preference for managing inboxes. For example, R1 trashed almost every bulk email after they opened and read it (see Table \mbox{\ref{tab:user_rate})}, \textit{``If it doesn't require actions, or not related to my work, I just trash it.''}. In contrast, R2 did not trash any email. 
    
    Trashing a bulk email \hlc[white]{did} not necessarily mean that it was useless. For example, R5 trashed a relevant bulk email after reading it: \textit{``I skimmed it because I was going to see whether we have all clear on the water issue ... I delete it because the water is good, no action needed.''}, and trashed a relevant bulk email without opening it: \textit{``I didn't need to open it because all the useful information are in the title --- come and get <person name> ... I trashed it.''}.

\end{enumerate}

% Open rate does not mean the same thing to recipients as it does to communicators: (a) we have learned they often open/delete rapidly, and (b) they did not see high open rate as inherently positive -- it may just mean the title did not tell them enough to confidently skip the email. Whether opening an email or say, whether clicking on an email to remove the ``unread'' tag, was more related with the population and the recipient's email usage preference instead of the number of emails received, see Figure~\ref{fig:detail_count}. Faculty opened more organizational emails compared to staff ($95.58\%$ vs $73.02\%$ in all organizational emails, $96.82\%$ vs $78.41\%$ in organizational non-bulk emails, $92.22 \%$ vs $51.82\%$ in organizational mass emails, $94.12\%$ vs $50.00\%$ in organizational news emails).

% \begin{figure}[!htbp]
% \centering
%   \includegraphics[width=0.65\columnwidth]{figures/detail_count}
%   \caption{Relationship between the open rates of organizational emails and the number of organizational emails received within a week, collected by letting recipients inputting searching commands for emails sent from the organization received in the past week in Gmail and counting resulting emails. See appendix for commands.}~\label{fig:detail_count}
% \end{figure}

Though ``open rate'' was a flawed metric, communicators had few tools in the bulk email system to get performance besides open/click rate: \textit{
    ``I don't know, how long people spend, whether they share with other people.'' (C1)
}

\noindent\textbf{Implication 1: Provide End-to-End Metrics.} Focusing on ``open rate'' distracted communicators from real measurements of both channel and bulk email's effectiveness. Communicators usually were not involved in the transactional process of the bulk emails. They did not know whether the communication really worked out --- the recipients read the emails and took corresponding actions to complete the organization's goals. Thus showing metrics on the transactional process of bulk emails to communicators could be helpful, for example, showing how many people missed the due dates of events that were sent to them in emails. When connecting heterogeneous systems is difficult, we could build systems similar to some human resources analytic systems \mbox{\cite{rasmussen2015learning}}, like collecting assessments from clients/leaders for each bulk email.

\vspace{-0.25in}

\begin{table}[!htbp]
\centering
\resizebox{0.85\textwidth}{!}{%
\begin{tabular}{|c|c|c|c|c|c|c|} 
\hline
\textbf{\#} & \textbf{type} & \textbf{\# investigated} & \textbf{\# opened} & \textbf{\# read in detail} & \textbf{\# trashed} & \textbf{\# opened and trashed}  \\ 
\hline
R1          & staff         & 20                           & 5                         & 3                                 & 14                         & 5                              \\ 
\hline
R2          & staff         & 23                           & 19                        & 12                                & 0                          & 0                              \\ 
\hline
R3          & staff         & 12                           & 4                         & 1                                 & 1                          & 1                              \\ 
\hline
R4          & staff         & 8                            & 5                         & 2                                 & 1                          & 0                              \\ 
\hline
R5          & staff         & 20                           & 15                        & 6                                 & 3                          & 2                              \\ 
\hline
R6          & faculty       & 17                           & 12                        & 1                                 & 7                          & 4                              \\ 
\hline
R7          & faculty       & 22                           & 17                        & 2                                 & 8                          & 5                              \\ 
\hline
R8          & faculty       & 24                           & 10                        & 2                                 & 3                          & 0                              \\ 
\hline
R9          & faculty       & 17                           & 12                        & 4                                 & 14                         & 9                              \\
\hline
\end{tabular}
}
\caption{The number of bulk emails investigated/opened/read in detail/trashed/opened and trashed. The data was self-reported by participants on the specific bulk emails we investigated during the interviews.}
~\label{tab:user_rate}
\end{table}

\subsubsection{ Diverse types of communication with different content and delivery channels, and different matchings between content and channels.}
~\\
\noindent\textbf{A. Bulk email's contents.} Communicators mentioned three types of bulk email's content:

\begin{itemize}
    \item Transactional content: which aimed to facilitate the work of the university, and asked recipients to take action. Communicator C4 talked about the goals of transactional bulk emails: \textit{
       ``Generally, we want you to change your behavior or be aware of something happening, or think differently about behavior or things like that.''
    }
    
   Some transactional content was urgent and required immediate actions:
    
    \begin{quote} \textit{
         Around something we want people to immediately know about, we usually send those transactional versus commercial, we have that option within the salesforce. (C1)
    }\end{quote}
    
    % Transactional subsections were accurate in targeting. Their goal is to reach right people:
    % \begin{quote} \textit{
    % For faculty who use canvas, we send them direct transactional emails. For example, it's the beginning of this semester you need to do these for your course. (C5).
    % }\end{quote}

    \item Highlighted News: These were the most important news that offices \hlc[white]{wanted} employees to be aware of (e.g., announcements by the university president of new officials), though for most employees these would not be actionable. A communicator introduced this kind of email:
    
    \begin{quote} \textit{
        The broadest thing we do is that we have a blog, that we produce content and stories every month, and every month we sent highlights of the blog, that we want people to know about, to be aware of. (C1)
    }\end{quote}
    
    \item Good-to-know News: These were other updates that offices felt employees would benefit from knowing, or that the institution would benefit from having more people aware of. Communicators \hlc[white]{recognized} that individual employees could miss them without consequence. 
    
    % Recipients may find some of these interesting and others not at all valuable.
    
    % Both highlighted news and good-to-know news are usually aimed to reach the large community of the university.
    % \begin{quote} \textit{
    %          News releases, research-related news are the things that we want to send to a large community. (C1)
    % }\end{quote}

\end{itemize}

A bulk email's content might be good-to-know for some recipients, while being transactional for some others. We asked a communicator whether all the recipients (all students, faculty, and staff) of a bulk email about campus-safety tips should read it: \textit{
   `` Students for sure. For faculty and staff, it's a reminder, but it's not something you have to know.'' (C5)}

\noindent\textbf{B. Channels and Matchings between Message' Content and Channels.}  There were 3 channels of bulk emails: single bulk emails (mass emails), newsletters (newsletter emails), and redistribution mechanisms (sent to the communicator's contacts in decentralized units, see 6.5.2). The matching between types of messages' content to types of channels were different across communicators. Some communicators sent good-to-know news through newsletters and some sent them through single bulk emails. Some communicators sent highlighted news and transactional content through single bulk emails and some communicators sent them through newsletters:
    \begin{quote} 
        \textit{We send single messages [individual bulk emails] when something is really timely we need to get it out, and when it’s critical that people receive and see it. If something is a story, less critical then that might find its way into our newsletter instead. (C1)}
        
        \textit{If there is an actionable thing we want people to do, that typically will go into a newsletter bulletin. The single message [individual bulk email] would be like a topic that needs to have a lot more detail but not necessarily an action needed to come out of it. So it’s an awareness thing. (C5)
    }\end{quote}
    
    % \item Some communicators sent good-to-know news through newsletters and some sent them through single OBEs. A communicator talked about the newsletters she sent:
    % \begin{quote} \textit{
    %     The newsletter is about updates and changes they does not specifically need to know. (C5) 
    % }\end{quote}
    
    % We asked the same communicator whether a single OBE that was sent to 2000 recipients was important to all of them:
    
    % \begin{quote} \textit{
    %     Some days some would, other days they might not. They are good-to-know messages --- here are where the university showed up in the media. (C5)
    % }\end{quote}

The different matchings between bulk email's content and channels between communicators resulted in good-to-know news, highlighted news, or transactional content being put into a newsletter at the same time. We asked C5 whether the recipients of an email should read all of the 10 messages in it: \textit{``
Not really. So we really want the message ``have the ID card with you'' get out, that was important, the most actionable thing, but the rest of it is to communicate that we care.'' (C5)
}

% \begin{quote} \textit{
% I break it down into different categories for which we have administrative units ... For different people the important part is different, they may not apply to all people, but as an administrator of the office, we encourage to. (C3)
% }\end{quote}

% \begin{quote} \textit{
% Not really. So we really want the message (``have the ID card with you'') get out, that was important, the most actionable thing, but the rest of it is to communicate that we care. (C5)
% }\end{quote}

Recipients needed to go through all the messages in each email to screen messages relevant to them. A recipient talked about a newsletter with over 30 messages:
\begin{quote} \textit{
    What I do is I go through it, ok, <program name> ... I'm not interested in it, upcoming programs ... if I see something interesting then I'll read and click on that. (R6)
}\end{quote}

This mismatch between bulk email's content and channels across different communicators caused confusion about what the newsletters were used for and whether the recipients should view them as important.\\

\subsubsection{Communicators' various roles --- Communicator Directors VS Communicator Staff.}
~\\
\noindent Communicators' worries with bulk emails varied based on their roles. When we asked what needed to be improved in the bulk email system, the communicator directors cared about the engagement of the community, \textit{``We struggle on the engagement, make people know the service we have.'' (C2)}, \textit{``Our hardest things are how to let people feel that we all have a part of responsibility in safety, each individual have to make effort, it's much easier for us to keep the campus safe.'' (C4)}

Communicator staff would like more tools in the bulk email system which could support their operational work, \textit{``It's great to have people to say "I don't want this type of thing" or "Please send more content like this."'' (C5), ``From a user perspective, I think that maybe adapting some common templates could be helpful.'' (C6)}

\subsection{Different Perspectives on Bulk Email's Values}
\subsubsection{Communicators' assessments --- high-level bulk email is important.} 
~\\  
High-level bulk emails were the emails communicating what was going on generally in the university, like changes in the benefits, leadership, public safety, administration, or working tools; they were usually from the university-wide offices (see Table \ref{tab:sender_level} for the definitions of emails levels). They were recognized as the bulk emails with general importance by communicators. Communicators usually sent them to all faculty and staff: \textit{
         ``Public safety is everybody who works at the U should know about, and any big change in benefits or leadership will affect everybody.'' (C1)
    }
    
Communicators viewed the bulk emails from the leaders important, as we asked C5 what kind of emails were considered more important: \textit{
        ``Message from the president, from the vice president of university services about campus safety, those types of things.'' (C5)\\
    }
% \subsubsection{Managers' assessments.} 
% Managers sometimes agreed with their employees (recipients) that some OBEs were not important, and sometimes disagreed with their employees. Within the 31 emails we let managers assessed, there were 5 emails that the recipient and the manager both agreed that they should open; and there were 13 emails that the recipient and the manager both agreed that they should left unread. There were 8 cases where employees opened these emails while their managers thought they should skip it. These 8 cases were interesting to employees, but not particularly job-related: 1 of them was about an employee's previous workplace at the university, 4 of them were about events and programs (alumni marketing, a company's on-campus recruitment feedback questionnaire, another college's new programs), 2 of them were about job openings that an employee passed along to other people, 1 of them is about child care planning.

\subsubsection{Managers' assessments --- have senses about high-level bulk emails.}
~\\
Managers sometimes agreed with their employees (recipients) that some bulk emails were not important. Within the 31 emails we let managers assessed, there were 5 emails that the recipient and the manager both agreed that they should scan; and there were 13 emails that the recipient and the manager both agreed that they should be left unread.

Managers sometimes disagreed with their employees. There were 8 cases where employees opened these emails while their managers thought they should skip it. These 8 cases were interesting to employees, but not particularly job-related, like alumni marketing, a company's on-campus recruitment feedback questionnaire, etc.

Managers thought that employees should have some sense about what was going on at the high level of this university, about leadership and administration changes, as the 5 cases shown in Table \mbox{\ref{tab:disagree}}. They were the bulk emails which managers thought their employees should read while the employees did not, or thought the employees should read in detail while they just scanned. For example, M2 thought R6 should read the bulk email with the title \textit{``Your Help Matters Now''} and content \textit{``Legislative session ends Monday. Please take 60 seconds to make sure that when the final negotiations are done.''} in detail, but R6 just left it unread.\\

\subsubsection{Employees' assessments --- high-level bulk email is irrelevant.} 
~\\
 Employees often skipped bulk emails about legislation of the university and leadership changes (see Table \ref{tab:disagree}) because 1) they had time pressure; 2) they felt these emails were too high level to be related to themselves as a non-leader staff: \textit{
    ``I skimmed and deleted it. It's a program that's important to my boss (the dean), but not directly related to my office.'' (R7)
    }
    
Actually, the higher the bulk email's level, the lower the probability that the bulk email would be opened/read in detail, as we showed in Figure \ref{fig:open_level}, where we compared the levels of the investigated bulk emails (see Table \ref{tab:sender_level}) with their open/read in detail rates. A bulk email from a university-wide office had 48\% chance of being opened and 14\% chance of being read in detail, while for a bulk email from a department's office these numbers would be 95\% and 57\%. 

The bias on high-level bulk emails made employees miss some useful emails. Employees actually found some of them helpful when being asked to read the unread bulk emails which the managers thought they should. R6 found that the \#1 bulk email in Table \mbox{\ref{tab:disagree}} should be read: \textit{``When I read it now, I may click on here and take 30 seconds to make sure the final negotiation. So I didn't realize that it was so easy to click this button. I was thinking it might involve more like writing a short letter. ''}

\begin{table}[!htbp]
\centering
\resizebox{\textwidth}{!}{%
\begin{tabular}{|p{0.2cm}|p{1.8cm}|p{1.6cm}|p{1.8cm}|p{4.5cm}|p{2cm}|p{5cm}|} 
\hline
\textbf{\#} & \textbf{Manager \& Recipient} & \textbf{Manager's Expectation} & \textbf{Employee's Reaction} & \textbf{Title}                                                                                            & \textbf{Sender}                  & \textbf{Employee's Reason}                                                                                       \\ 
\hline
1           & M2 - R6                       & Read in Detail                 & Unread                     & Your Help Matters Now                                                                                     & University Government Relations  & I just deleted it because I think it's important to advocate for the university but I don't do that for myself.  \\ 
\hline
2           & M2 - R6                       & Scan                           & Unread                     & <College Name>'s Update • Summer 2019                                                                        & Newsletter Center of <College Name> & I was too busy, this email is not like so important.                                                             \\ 
\hline
3           & M1 - R5                       & Scan                           & Unread                     & Reminder - Incorporating GoldPASS into your Career Courses session tomorrow                                & University Employer Relations    & I will not open it, because it's for undergrad department.                                                       \\ 
\hline
4           & M1 - R5                       & Scan                           & Unread                     & PeopleSoft Navigation Update: MyU  PeopleSoft Outage This Weekend                                         & PeopleSoft Update Team           & I already learn it from other sources.                                                                           \\ 
\hline
5           & M2 - R6                       & Read in Detail                 & Scan                         & {[}<College Name> - Staff] <College Name> Highlights, Leadership Announcements and Resources to Share, 05/20/19 & <College Name> Leader's Office      & I am busy and I will spend less time on this.                                                                    \\
\hline
\end{tabular}}
\caption{Title/sender of the bulk emails which managers thought their employees should read while they did not, or thought the employees should read in detail while they just scanned,  and the employees' reasons.}
~\label{tab:disagree}\vspace{-0.3in}
\end{table}

\begin{table}[!htbp]
\begin{minipage}[b]{0.32\linewidth}
\centering
    % \begin{table}[
    % \centering
    \small{
    \begin{tabular}{|p{0.2cm}|p{1.7cm}|p{1.7cm}|} 
    \hline
     \textbf{\#}  & \textbf{Level}  & \textbf{Meaning}                                                               \\ 
    \hline
    1             & University      & Sent from university wide offices.  \\ 
    \hline
    2             & College         & Sent from college wide offices.     \\ 
    \hline
    3             & Department      & Sent from department wide offices.  \\
    \hline
    \end{tabular}}
    \caption{The definition of email's levels. For example, ``university-wide'' offices indicates that the office was in charge of sending bulk emails to recipients across the university.}
    \label{tab:sender_level}
    % \end{table}
\end{minipage}\hfill
\begin{minipage}[b]{0.65\linewidth}
\centering
% \begin{figure}
    \centering
    % \begin{figure}
      \includegraphics[width=0.9\columnwidth]{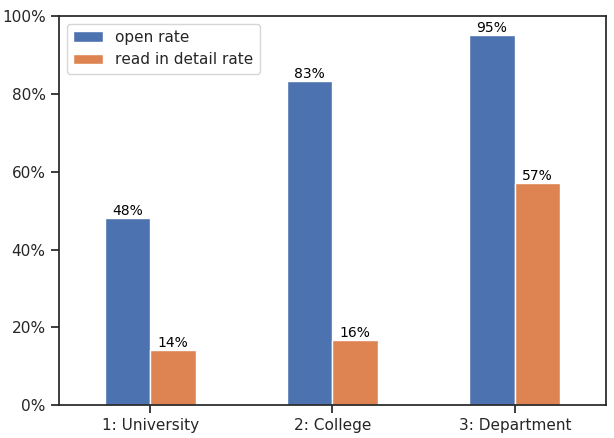}
      \captionof{figure}{The self-report open/read in detail rate of bulk emails of different levels collected in the interviews with recipients.}~\label{fig:open_level}
% \end{figure}
\end{minipage}
\end{table}

\subsection{Costs of Communication}

\subsubsection{Recipients’ feeling of burden and their reading strategies.}
~\\
Controlling the burden on recipients was not one of the goals of the current bulk email system, as stated by a communicator when asked whether a recipient's time cost was considered: \textit{``No, the email we sent are most not nice-to-know emails. They are required information for people.'' (C2)}

Receiving lots of emails did not necessarily make recipients feel a burden. R9 received over 774 emails per week but still did not get a sense of email overload. R9 talked about their strategy: \textit{``I can handle it (bulk email) ...  I scan it right away ... I delete the things I know I don't have to read ... I look at the subject ... Mentally you do it in hierarchy --- you determine what's the next, what's the first.''}

% R2 only received 188 email per week but  (see Table \mbox{\ref{tab:email_number}}), and felt a burden from bulk emails.}

The reading strategy "checking the content whenever a new email came in" might bring on a sense of burden, like R6 said: \textit{`` I checked my email pretty quickly. Even when I am working on something, I may go back to my email to see is there anything important that's coming, and I respond pretty quickly ... To be honest, sometimes I felt overwhelmed.''}. This result matched with Mark's findings on work emails and productivity \mbox{\cite{mark2016email}}.
 
Retrieving the right bulk email from inboxes could be a burden: \textit{``I might miss something. Sometimes my supervisor ask me whether you received it --- I trashed it then I try to find it. That's frustrating because in Gmail sometimes I can't find it. You try to think what's in the subject line, date ...\mbox{''} (R4)}.

% The recipients who trashed OBEs frequently tended to feel a lighter burden. For example, as shown in Table \ref{tab:user_rate}, R1 almost trashed every OBE: \textit{``I can handle it because I just don't read them.''}

\noindent\textbf{Implication 2: Make Recipients' Burden Visible to Communicators.} The time cost caused by bulk email and the collective burden put on recipients was inapproachable for the communicators. They did not know whether they sent a bulk email to too many recipients and let them spend unnecessary time, or collectively let some recipients feel overloaded. For example, we could monitor employees' time spent on bulk emails, or the financial cost of each bulk email \cite{jackson2006simple}.\\

\subsubsection{Communicators' Costs --- Loss of channel credibility (organizational communication capital).}
~\\
% Communicators wanted to maintain channel's credibility. They limited the use of channels.
% \begin{quote} \textit{
%     We want to be careful to not overuse the safeU system. There is very clear time when they are allowed to use that. (C4)
% }\end{quote}
% However, communicators' clients specified certain audiences such that they needed to send some emails
Communicators, requested by their clients, had to send some bulk emails at the cost of the sender's credibility. We asked a communicator whether the credibility of bulk email channel was considered:
\begin{quote} \textit{
 Yes, we definitely think about it (the credibility of the sender's name). But we work with different clients, they request the audience they want. (C1)
}\end{quote}

% Recipients learned from experience and stopped opening emails from low-credibility senders, or reading them in detail. 4 out of 9 participants mentioned that they used to not open emails from some senders because they believed that the OBEs from those senders were uninteresting to them. 
% For example, R6 trashed an email from university alumni association advertising football tickets sales:
%
% \begin{quote} 
% \textit{I recognized the sender and most times I deleted it right away. So I think again. I am not a (a sports team) fan so I wasn't so interested in it. (R6)
% }\end{quote}
% R8 did not open an email from the sender ``XXX Funding Opportunities'':
% \begin{quote} \textit{
%  Because we get this every month. From my past experience, they are just information from Federal Grant Agencies. It's not very useful --- too general, not very related to me. (R8)
% }\end{quote}
Recipients learned from experience and stopped opening bulk emails from low-credibility senders. For example, R9 trashed an email from the university fitness center without opening it: \textit{
``I recognized the sender and trashed it because I am not interested in it.''
}

\noindent\textbf{Implication 3: Make the Credibility of Bulk Email Channels Visible to Communicators.} There was no tool to tell communicators how credible a channel is to its recipients --- instead they discarded ineffective channels and replaced with new channels, which was a time consuming strategy. Thus quantifying channels' credibility and setting standards of usage of channels could be helpful. For example, calculating the credibility and success probability of communication \mbox{\cite{dewatripont2005modes}}, and avoiding sending optional bulk emails via high-credibility channels.

% R9 left an email with title \textit{``Can you help the University of XXX at the Legislature''} unread:
% \begin{quote} \textit{
%  I did not read it because I know what he is going to say. Basically, the article asked you to contact your legislators to support a capital request. (R9)
% }\end{quote}
% On another side, recipients spent a lot of time in dealing with  bulk emails each week. In our survey with 162 responses (see Figure \ref{fig:bulk_time}, 95 participants spent less than 10 minutes in reading bulk email every day, 62 participants spent 10~30 minutes every day, and 5 participants spent more than 30 minutes. From the view of an institution with over 20,000 employees, the institution will spend more than 1379 hours in reading bulk emails every day, which is a huge cost in time and money.

% \begin{figure}
% \centering
%   \includegraphics[width=0.6\columnwidth]{figures/bulk_email_time}
%   \caption{The Time Spent on Reading Bulk Emails.}\vspace{-30pt}~\label{fig:bulk_time}
% \end{figure}
\subsection{Understanding Current Practice and Its Failings}
\subsubsection{ Lack of personalization tools.}
% \subsubsection{Email Design}
% ~\\
% \noindent\textbf{A. Let recipients find key subsections by themselves.}
~\\
 Communicators tried to help recipients find important messages in a newsletter by putting important messages in the top position or putting their keywords in the subjects: \textit{``Sometimes we put important messages in the subject line. It's not the same importance (to all recipients), but they can recognize that it's important or not. We do try to keep it relevant and make it clear at the top.'' (C1)}
 
 However, as C1 said, importance of the messages in a bulk email may vary for different recipients. Recipients tended to close a newsletter whose first message was not relevant; and sometimes they did miss the later messages which were useful to them. R6 told us why a bulk email was unread and R6 found it useful when we asked they read the email:
 \begin{quote}\textit{
    They started with sports, then I thought it's not relevant to me. But when I go back with you, I found out that it has research that I am interested in. (R6)
    }\end{quote}
    
Given this situation, communicators thought that personalization of email designs would be helpful, but they did not have the technology to do personalization. Currently, all recipients receive the same subject lines/subtitles/content for each bulk email. We asked a communicator whether they had felt the need for personalization of emails):
\begin{quote} \textit{
     We haven't, although it's something we would like to explore doing. I don't think our team has technical expertise. I don't know where and how to get that type of thing. (C6)
}\end{quote}

\noindent\textbf{Implication 4: Explore Personalization.} Develop personalization tools for communicators to design bulk emails. For example, for a user, estimate the relevance of each message in a bulk email to this user and rank the messages from high to low relevance \cite{trespalacios2016effects}.\\

\subsubsection{Flawed Email Targeting Tools}
~\\
\noindent\textbf{A. Target through mailing lists --- recipients cannot opt-out.} There were some mailing lists built for certain groups, e.g. all faculty of the university, all staff of a department. Communicators assumed that recipients would unsubscribe from mailing lists they were not interested in. Communicators also designated certain lists without opt-out option to avoid core recipients opt-out. For example, a communicator talked about a monthly newsletter:
\begin{quote} \textit{
    For the people who work at administrators, they cannot opt-out of that. I check this. Everyone else is optional, they can unsubscribe. (C3)
}\end{quote}

% A communicator talked about a monthly newsletter about highlights of a blog of the president, and research-related news:
% \begin{quote} \textit{
%     Those are subscribers, internal and external. We send that through mail channel once a month. We have another newsletter called ``insight of vpr'' specific to vice president research's staff. You can't subscribe to that --- that's more specific to internal people. (C1)
% }\end{quote}

However, there were cases that a mailing list was used to send important and unimportant bulk emails at different times. Recipients could not opt-out from this mailing list even though unimportant bulk emails were sent through it because they would not then receive the important emails sent through that mailing list in the future. For example, a communicator talked about an optional bulk email sent to communication directors:

\begin{quote} \textit{
It's optional for them to take action to do what I ask them to do. This is meant to be kind of helpful service I'm giving ... They could (opt-out of this mailing list) but this is a mailing list for the communication directors so they probably won't. (C5)
}\end{quote}

% Based on the assumption that recipients would opt out mailing lists if they were not interested in them, recipients were put on mailing lists on default by communicators. Recipients did not know when or why they were put into some mailing lists and were unsure whether they could opt out these mailing lists, worrying that these emails could contain important information in the future, even when the OBEs they received from these mailing lists were irrelevant to them. A recipient received an email from university fitness center. We asked the recipient did she know why she received this email and did she think about unsubscribe from it:
% \begin{quote} \textit{
%     I thought it sent to everybody. Though it said it sent to me, but I am sure me is like the thousands in the university. This kind of things is expected to be sent to a faculty, but I never took a class there. I should (unsubscribe it), I know. But you just never know. After you unsubscribe it, you may need it. It could be important. This may not be important to me now, but 10 years later, it may be important. (R9)
% }\end{quote}

\noindent\textbf{B. Target through querying position-database --- limited to the human resource system.} If there were no mailing lists existing for the target population, communicators would query the database of employees. However, the current bulk email system only connected with the human resource system, which only supported querying based on positions, but not on recipients' networks or interests. A communicator introduced this process:

\begin{quote} \textit{
 We will go to HR, and ask them to pull a list from PeopleSoft. Sometimes, for example, we were trying to reach everyone with family, so we go to HR, and they were not able to pull out. They can only do if the data is in PeopleSoft and we can get it by job code. (C5)
}\end{quote}

A communicator talked about a bulk email whose target population was all employees who had not downloaded documents from Webex. But the communicator did not have such information in the bulk email system, it was sent to all Webex users instead:

\begin{quote} \textit{
    This is the final reminder to Webex users: action required to download reports. Receivers are people who have been using Webex in the last year. That went to 9000 receivers.} 
    
    \textit{We cannot tell who has downloaded their file or not. The audience is who is using Webex and have files. Anybody who has already moved to Zoom, downloaded their files don't need to take action, but we don't know who are them. (C2)
}\end{quote}

\noindent\textbf{C. Redistribution mechanisms --- no tracking of bulk emails.} If communicators did not have enough information to target recipients, expecting that lower-level units would have more information about their employees, communicators sometimes sent bulk emails to their contacts in units and asked them to forward these emails to the related recipients:

\begin{quote} \textit{
     We have in each unit one person in charge of the facility things. And they may have people in their department, in their buildings. Sometimes we will send information to the unit facility leads, which is a small set of people (author's note: this is a set of 400 to 500 people). That's when we want other people to distribute something more locally versus from the University Relations. We want people in <college> to get some messages [emails] from their unit. We may ask them to distribute that message [email]. (C4)
}\end{quote}

However, communicators could not track whether and how the contacts sent these bulk emails, nor the performance of these bulk emails, when we asked a communicator if they could track the performance of the bulk emails sent to the contacts: \textit{``Unfortunately we can not track that.'' (C6)}

% We asked a communicator could she track the performance of the OBEs sent through the distribution mechanism:

% \begin{quote} \textit{
%     Unfortunately we can not track that. So every month we send them an email saying your dashboard shows how well you are doing with I-9s out there, please go to look at it, so we kind of do reminder. (C6)
% }\end{quote}

\noindent\textbf{D. When it is hard to target --- overwhelm everyone.} As both mailing lists and querying were limited to certain scenarios, precise targeting was difficult for the communicators. C6 introduced a group difficult to narrow: \textit{``Open enrollment is tough because we don't know how many people need to take action. That's kind of a dilemma. Because we try to make that system easier for people so you don't have to do anything. But that makes it harder for us to do communications.''}

Motivated by ``getting the proof of delivered'', communicators thought that it was better to overwhelm everyone than to miss a single person. The bulk email above finally went to all employees: \textit{``We moved everyone to do it to ensure that no one is left behind.'' (C6)}

% To ensure the core population receives these OBEs, communicators usually sent OBEs to a large community, much wider than the core targeting groups of the OBEs. We asked a communicator that were there groups that were just too hard to narrow such that they needed to send to all people:

% \begin{quote} \textit{
%     Probably, for example, with TechComm, we have a list of people we send information about start-up workshops that are happening. That's very targeted to specific colleagues, department, at different campuses. (C1)
% }\end{quote}

% Some emails were sent to all staff and faculty of the university. We asked a communicator who were the core recipients of a daily newsletter that was sent to all staff and faculty:
% \begin{quote} \textit{
% University leadership, faculty, and staff who indicated their interests. It's hard for us to know that. (C5)
% }\end{quote}

% Communicators thought that it was better to overwhelm everyone than to miss a single person:
% \begin{quote} \textit{
%     Open enrollment is tough because we don't know how many people need to take action. That's kind of a dilemma. Because we try to make that system easier for people so you don't have to do anything. But that makes it harder for us to do communications, so we moved everyone to do it to ensure that no one is left behind. (C6)
% }\end{quote}

``Sent to all'' was one of the reasons that recipients received irrelevant bulk emails. R4 talked about a bulk email they found not useful: \textit{``This is what I call blanket message [email] --- it goes to everyone, it's common knowledge, it's not that related to me.''}

\noindent\textbf{Implication 5:} Precise targeting is difficult. As previous research showed that emails' labels would change recipients' behaviors \mbox{\cite{reeves2008marketplace}}, a mechanism could be designed to hint an email's value to its recipients before they opened it, for example, labeling whether a bulk email is optional or mandatory, action-needed or not \cite{10.1145/1978942.1979456}. \\

% A consequence caused by sending emails wider than the core population is that the recipients often receive emails unrelated to them. A recipient received every update email from a google group built for guiding employees to change reimbursement software:

% % In our interview, we asked the recipients to give ratings on the relevance of bulk emails they received from 1 (least relevant) to 5 (the most relevant). Over 154 emails we collected, $54.1\%$ of them had low relevance (rating was 1 or 2).
% \begin{quote} \textit{
% This is just a broad discussion. So people have questions, they sent it here. So if I have questions I sent it to you then this might be relevant, but I am just on the mailing list so I get all the emails about it. (R3)
% }\end{quote}

% As a result, some recipients said that they never open emails from this office ...

% \textbf{Various Audience}

% Each communicator served various audience. They can't access detailed or up-to-date information about the audience.

% Sometimes it's difficult to narrow down a mailing list by a bulk emails' optimal targeting population because of lacking of information.

% \subsection{Results --- Evidence of Failure: Special Handling of ``Important OBEs''}
% The current OBES is viewed as inadequate for ensuring that important information are reliably received and read; therefore communicators have developed a set of special treatments to overcome the shortcomings of standard OBEs delivery. \\

\subsubsection{Communicator transferred the responsibility of being aware}
~\\
The responsibility of knowing a bulk email was transferred to the recipients after the communicators sent it out. A communicator talked about a newsletter from the college's dean that they sent to all employees weekly, with 17 messages and 5 tweets from the dean:

\begin{quote} \textit{
They are all important emails, sent once a week. So our expectation is --- ``This is the important email you get from your college, regarding your job. This email is the business of the college, you should read it.'' (C3)}\end{quote}

However, recipients found the bulk emails were too many or too long for them to filter out unimportant ones, find important ones, and be aware of all of the content. They felt that it was not the recipient's responsibility to be aware of all the content. A recipient talked about a weekly newsletter sent to all employees about government activities:

\begin{quote} \textit{
It took so much time to read, 10 paragraphs, nobody gonna read that, it's gambling. People use emails like a safety cover, but there is no understanding from the recipients' side. The responsibility shifts from the sender to the recipient, that's unfair. (R3)}\\\end{quote}

\subsubsection{Vicious cycle --- recipients did not read then communicators sent more.}
~\\
Similar to Randall's findings in organizational communication \mbox{\cite{SCHULER1979268}}, the ineffectiveness of bulk email system also brought a vicious cycle --- 1) recipients did not read bulk emails on time; 2) to get recipients to read bulk emails/take actions on time, communicators sent bulk emails multiple times/more widely; 3) recipients received more irrelevant bulk emails, lost trust in bulk email channels, and read less bulk emails in the future. A communicator talked about the collection process of the employee engagement survey: \textit{``Those messages [emails] start from September, October, in terms of emails, go through January ... we do have to do emails more than once because our email open rate is about 61 or 62 percent.'' (C6)
}

Communicators did not realize the existence of the cycle and thought repeated bulk emails would be appreciated: \textit{
``My assumption is if somebody is emailing me about a service that is going away, and I have taken actions, I will never need to open it, but I will be grateful if I get reminders.'' (C2)}

However, recipients found they usually opened a bulk email and found that they already read it/still not interested/already took action. A recipient trashed a bulk email about reimbursement:

\begin{quote} \textit{
It is important and relevant but this is not the first time she sent. I already picked up what I need ... I only read the first email a couple of months ago ... She even attached the same document again ... so I just deleted it. (R9)}\end{quote}

\noindent \textbf{Implication 6: Provide Feedback to Communicators.} To interrupt the vicious cycle, recipients' feedback could be collected, such as whether the recipient wanted more like this/less like this (like what was applied in social medias \cite{khan2017social}), how much time the recipient had spent on it, whether they had already seen/replied to this bulk email. Communicators then could decide whether to keep sending these bulk emails to the recipient, or only distribute reminders to people that have not read or confirmed them.

\section{Discussion}

% 3) the practice was inconsistent --- with limited personalization tools, communicators tried several methods of designing and distributing bulk emails; however, recipients found these practices inconsistent between different communicators; 4) there was a vicious cycle --- recipients refused to read bulk emails, then communicators tried to send more and therefore lower the credibility of bulk email channels.

\subsection{Align Different Stakeholders' Priorities}
We found a large gap between different stakeholders' priorities. The communicators thought they were sending important high-level information about the organization; and communicators' clients wanted their messages to be sent broadly. However, our recipients naturally did not view high-level bulk emails as important. This gap was caused by different stakeholder priorities.

On one side, the major task for communicators was to get the ``proof of delivering'' of bulk emails to their clients, whose first priority was to let the target population be aware of their information and take appropriate actions. On the other side, recipients, who were collectively burdened by the communicators, wanted to receive bulk emails that were interesting and relevant to them.

Outside the organizational environment, these priorities were helpful for companies, like sending advertising emails to customers broadly \mbox{\cite{7004277}}. However, within the organization where all employees had responsibilities of completing the organization's tasks and minimizing the organization's costs, this misalignment of priorities brought costs to the organization. The communication channels' credibility was harmed. The organization's tasks were not done.

Thus it is important to align different stakeholders' priorities, and also to include the priority of the organization itself in the bulk email system. Recipients should be reminded that besides the interesting stuff, they have the responsibility of being aware of what is going on in the organization. Communicators and their clients should be reminded that besides distributing their information, they have the responsibility of maintaining the effectiveness of organizational communication channels and avoid employees having to spend time on unnecessary emails. At an extreme, a Key Performance Indicator System \mbox{\cite{kueng2000process}}, which takes the organization's priority into consideration, could be integrated into the bulk email system. For communicators, there could be indicators on the time cost, the credibility cost, and their clients' feedback for each bulk email. For recipients, there could be indicators on the speed of taking actions required by bulk emails, the time spent on reading the bulk emails important to the organization.

\subsection{Limitations --- Organizational Culture/Structure's Influence on Bulk Email System}
This is a qualitative case study of one study site. The observations might not be generalizable to all kinds of organizations. It is possible that the stakeholders of other organizations would have different practices/perspectives on bulk email systems with respect to their organizations' cultures/structures.

The organization we studied had a hierarchical management structure --- from university leadership (Board of Regents, Office of the President, Office of the Provost), to the college office (deans, academic affairs), to the department office (heads, program directors). The higher the level, the more information they had. Thus the information flow of the bulk email system was mostly one directional --- low-level employees got information from the high-level leaders by bulk emails, while high-level leaders rarely needed to get information from low-level employees by bulk emails.

As for stakeholders within an organization, bidirectional interaction proved to be an important factor in their engagement levels of organizational affairs \mbox{\cite{norris2017stakeholders}}. Thus this bulk email system's audience might have less willingness to participate in the bulk email communication and that might cause its ineffectiveness. It is possible that the recipients in the organizations who had a flat structure would have a higher willingness to participate in bulk email communication.

Though hierarchical, the organization's communication system was to some extent decentralized --- each communication offices worked for their own units, had their own goals, served their own clients from the same units or other units. There were no central alignments of these communication offices and some information would be sent multiple times from different offices --- these repeated bulk emails then became a burden that multiple communication offices collectively placed on recipients. As the previous research found, \textit{``decentralized organization is not a very efficient organization
model for work that requires a lot of intensive interaction
between different employees.''} \mbox{\cite{maki2004communication}}, it is possible that a bulk email system with a central communication office to arrange different communication offices' tasks and collaboration would work more efficiently.

\section{Conclusion and Future Work}
\noindent\textbf{Conclusion:} Through this study we found that the failure of the organizational bulk email system was systematic --- the organizational bulk email system had many stakeholders, but none of them necessarily had a global view of the system or the impacts of their own actions. First, the communicators' high-level clients wanted their messages sent broadly. Second, the communicators felt they were sending important high-level information through bulk emails; they got high ``open rates'' and only reported that to their clients. Third, the recipients viewed most of the bulk emails not relevant; they often opened and then rapidly discarded bulk emails without reading the details. Last, though the communicators across the organization worked together in a network that attempted to improve the quality of communication practice, different communicators used bulk email channels differently and there was a limited set of targeting and feedback tools to support their task.

In theory, the answer is simple; there is a long history of information filtering technology to help recipients avoid spam based on their preferences, and email marketing technology to help communicators advertise products to external recipients. But in an organizational setting, recipients or communicators' preferences are not enough; the organization's priorities matter. Recipients may not want to know certain things, but their employer expects them to know them nonetheless; communicators and their clients may want to send bulk emails broadly, but they need to consider the cost to the organization.

\noindent\textbf{Impact:} We hope that by this case study, we could 1) provide detailed information in how an organizational bulk email system works and fails, with respect to this organization's structure and nature; 2) to the best of our knowledge, this is the first work focusing on multi-stakeholders' perspectives on bulk emails within an organization; we hope our protocols on inbox artifact-walkthroughs are useful for future research; 3) given that many conventional in-person communication channels are no longer available for organizations in the current pandemic, organizational bulk email system now takes on even greater importance, especially in internal crisis communication; we hope our study provides facts and possible directions for studying organizational bulk email system in the remote-work environment.

\noindent\textbf{Future Work:} The next steps include: 1) collaborate with more organizations to study whether our findings in this case study are common; 2) informed by the design recommendations we make above, design an organizational bulk email system to support multi-stakeholder prioritization.  It would include personalization, but tied to both the interests of the recipient and the needs of the organization.  It would help communicators predict how their bulk emails would be received based on aggregated prior feedback and would provide mechanisms for managing the cost and value of bulk emails.  At an extreme, an organization might even allocate communication budgets to units to reduce wasteful or poorly-targeted organizational bulk emails, but even without this step, we believe the visibility could provide sufficient incentive to already well-motivated communicators.

\begin{acks}
This work was supported by the National Science Foundation (NSF) under Award No. IIS-2001851, IIS-2000782, and CNS-1952085. The authors would like to thank Anjali Srivastava, University of Minnesota GroupLens Lab members, and our reviewers for providing us with very valuable feedback.
\end{acks}

\bibliographystyle{ACM-Reference-Format}
\bibliography{sample-base}


\begin{thebibliography}{61}

%%% the Web address.

\ifx \showCODEN    \undefined \def \showCODEN     #1{\unskip}     \fi
\ifx \showDOI      \undefined \def \showDOI       #1{#1}\fi
\ifx \showISBNx    \undefined \def \showISBNx     #1{\unskip}     \fi
\ifx \showISBNxiii \undefined \def \showISBNxiii  #1{\unskip}     \fi
\ifx \showISSN     \undefined \def \showISSN      #1{\unskip}     \fi
\ifx \showLCCN     \undefined \def \showLCCN      #1{\unskip}     \fi
\ifx \shownote     \undefined \def \shownote      #1{#1}          \fi
\ifx \showarticletitle \undefined \def \showarticletitle #1{#1}   \fi
\ifx \showURL      \undefined \def \showURL       {\relax}        \fi
% invisible to TeX
\providecommand\bibfield[2]{#2}
\providecommand\bibinfo[2]{#2}
\providecommand\natexlab[1]{#1}
\providecommand\showeprint[2][]{arXiv:#2}

\bibitem[\protect\citeauthoryear{Al-Jarrah, Khater, and Al-Duwairi}{Al-Jarrah
  et~al\mbox{.}}{2012}]%
        {al2012identifying}
\bibfield{author}{\bibinfo{person}{Omar Al-Jarrah}, \bibinfo{person}{Ismail
  Khater}, {and} \bibinfo{person}{Basheer Al-Duwairi}.}
  \bibinfo{year}{2012}\natexlab{}.
\newblock \showarticletitle{Identifying potentially useful email header
  features for email spam filtering}. In \bibinfo{booktitle}{\emph{The sixth
  international conference on digital society (ICDS)}},
  Vol.~\bibinfo{volume}{30}. \bibinfo{pages}{140}.
\newblock


\bibitem[\protect\citeauthoryear{Alrashed, Lee, Bailey, Lin, Shokouhi, and
  Dumais}{Alrashed et~al\mbox{.}}{2019}]%
        {alrashed2019evaluating}
\bibfield{author}{\bibinfo{person}{Tarfah Alrashed}, \bibinfo{person}{Chia-Jung
  Lee}, \bibinfo{person}{Peter Bailey}, \bibinfo{person}{Christopher Lin},
  \bibinfo{person}{Milad Shokouhi}, {and} \bibinfo{person}{Susan Dumais}.}
  \bibinfo{year}{2019}\natexlab{}.
\newblock \showarticletitle{Evaluating User Actions as a Proxy for Email
  Significance}. In \bibinfo{booktitle}{\emph{The World Wide Web Conference}}.
  \bibinfo{pages}{26--36}.
\newblock


\bibitem[\protect\citeauthoryear{Andriof, Rahman, Waddock, and Husted}{Andriof
  et~al\mbox{.}}{2002}]%
        {andriof2002introduction}
\bibfield{author}{\bibinfo{person}{Jorg Andriof},
  \bibinfo{person}{Sandra~Sutherland Rahman}, \bibinfo{person}{Sandra Waddock},
  {and} \bibinfo{person}{Bryan Husted}.} \bibinfo{year}{2002}\natexlab{}.
\newblock \showarticletitle{Introduction: JCC theme issue: Stakeholder
  responsibility}.
\newblock \bibinfo{journal}{\emph{The Journal of Corporate Citizenship}}
  (\bibinfo{year}{2002}), \bibinfo{pages}{16--19}.
\newblock


\bibitem[\protect\citeauthoryear{{Balakrishnan} and {Parekh}}{{Balakrishnan}
  and {Parekh}}{2014}]%
        {7004277}
\bibfield{author}{\bibinfo{person}{R. {Balakrishnan}} {and} \bibinfo{person}{R.
  {Parekh}}.} \bibinfo{year}{2014}\natexlab{}.
\newblock \showarticletitle{Learning to predict subject-line opens for
  large-scale email marketing}. In \bibinfo{booktitle}{\emph{2014 IEEE
  International Conference on Big Data (Big Data)}}. \bibinfo{pages}{579--584}.
\newblock


\bibitem[\protect\citeauthoryear{Bekkerman}{Bekkerman}{2004}]%
        {bekkerman2004automatic}
\bibfield{author}{\bibinfo{person}{Ron Bekkerman}.}
  \bibinfo{year}{2004}\natexlab{}.
\newblock \showarticletitle{Automatic categorization of email into folders:
  Benchmark experiments on Enron and SRI corpora}.
\newblock  (\bibinfo{year}{2004}).
\newblock


\bibitem[\protect\citeauthoryear{Bellotti, Ducheneaut, Howard, Smith, and
  Grinter}{Bellotti et~al\mbox{.}}{2005}]%
        {bellotti2005quality}
\bibfield{author}{\bibinfo{person}{Victoria Bellotti}, \bibinfo{person}{Nicolas
  Ducheneaut}, \bibinfo{person}{Mark Howard}, \bibinfo{person}{Ian Smith},
  {and} \bibinfo{person}{Rebecca~E Grinter}.} \bibinfo{year}{2005}\natexlab{}.
\newblock \showarticletitle{Quality versus quantity: E-mail-centric task
  management and its relation with overload}.
\newblock \bibinfo{journal}{\emph{Human--Computer Interaction}}
  \bibinfo{volume}{20}, \bibinfo{number}{1-2} (\bibinfo{year}{2005}),
  \bibinfo{pages}{89--138}.
\newblock


\bibitem[\protect\citeauthoryear{Bermejo, G{\'a}mez, and Puerta}{Bermejo
  et~al\mbox{.}}{2011}]%
        {bermejo2011improving}
\bibfield{author}{\bibinfo{person}{Pablo Bermejo}, \bibinfo{person}{Jose~A
  G{\'a}mez}, {and} \bibinfo{person}{Jose~M Puerta}.}
  \bibinfo{year}{2011}\natexlab{}.
\newblock \showarticletitle{Improving the performance of Naive Bayes
  multinomial in e-mail foldering by introducing distribution-based balance of
  datasets}.
\newblock \bibinfo{journal}{\emph{Expert Systems with Applications}}
  \bibinfo{volume}{38}, \bibinfo{number}{3} (\bibinfo{year}{2011}),
  \bibinfo{pages}{2072--2080}.
\newblock


\bibitem[\protect\citeauthoryear{Beyer and Holtzblatt}{Beyer and
  Holtzblatt}{1999}]%
        {beyer1999contextual}
\bibfield{author}{\bibinfo{person}{Hugh Beyer} {and} \bibinfo{person}{Karen
  Holtzblatt}.} \bibinfo{year}{1999}\natexlab{}.
\newblock \showarticletitle{Contextual design}.
\newblock \bibinfo{journal}{\emph{interactions}} \bibinfo{volume}{6},
  \bibinfo{number}{1} (\bibinfo{year}{1999}), \bibinfo{pages}{32--42}.
\newblock


\bibitem[\protect\citeauthoryear{Buchanan}{Buchanan}{2012}]%
        {buchanan2012case}
\bibfield{author}{\bibinfo{person}{David~A Buchanan}.}
  \bibinfo{year}{2012}\natexlab{}.
\newblock \showarticletitle{Case studies in organizational research}.
\newblock \bibinfo{journal}{\emph{Qualitative organizational research}}
  (\bibinfo{year}{2012}), \bibinfo{pages}{351--370}.
\newblock


\bibitem[\protect\citeauthoryear{Carter, Donovan, and Jalleh}{Carter
  et~al\mbox{.}}{2011}]%
        {carter2011using}
\bibfield{author}{\bibinfo{person}{Owen~BJ Carter}, \bibinfo{person}{Robert
  Donovan}, {and} \bibinfo{person}{Geoffrey Jalleh}.}
  \bibinfo{year}{2011}\natexlab{}.
\newblock \showarticletitle{Using viral e-mails to distribute tobacco control
  advertisements: an experimental investigation}.
\newblock \bibinfo{journal}{\emph{Journal of health communication}}
  \bibinfo{volume}{16}, \bibinfo{number}{7} (\bibinfo{year}{2011}),
  \bibinfo{pages}{698--707}.
\newblock


\bibitem[\protect\citeauthoryear{Celik and {\"O}l{\c{c}}er}{Celik and
  {\"O}l{\c{c}}er}{2018}]%
        {celik2018importance}
\bibfield{author}{\bibinfo{person}{B{\"u}nyamin Celik} {and}
  \bibinfo{person}{Zeynep {\"O}l{\c{c}}er}.} \bibinfo{year}{2018}\natexlab{}.
\newblock \showarticletitle{The Importance of Corporate Email Service and
  Google Education Service for Education Institutions (A Case Study of Ronaki
  Duhok Education Company in Iraq)}.
\newblock \bibinfo{journal}{\emph{International Journal of Social Sciences \&
  Educational Studies}} \bibinfo{volume}{5}, \bibinfo{number}{1}
  (\bibinfo{year}{2018}), \bibinfo{pages}{168}.
\newblock


\bibitem[\protect\citeauthoryear{{Chakrabarty} and {Roy}}{{Chakrabarty} and
  {Roy}}{2014}]%
        {6970931}
\bibfield{author}{\bibinfo{person}{A. {Chakrabarty}} {and} \bibinfo{person}{S.
  {Roy}}.} \bibinfo{year}{2014}\natexlab{}.
\newblock \showarticletitle{An optimized k-NN classifier based on minimum
  spanning tree for email filtering}. In \bibinfo{booktitle}{\emph{2014 2nd
  International Conference on Business and Information Management (ICBIM)}}.
  \bibinfo{pages}{47--52}.
\newblock


\bibitem[\protect\citeauthoryear{Charmaz}{Charmaz}{2014}]%
        {charmaz2014constructing}
\bibfield{author}{\bibinfo{person}{Kathy Charmaz}.}
  \bibinfo{year}{2014}\natexlab{}.
\newblock \bibinfo{booktitle}{\emph{Constructing grounded theory}}.
\newblock \bibinfo{publisher}{Sage}.
\newblock


\bibitem[\protect\citeauthoryear{Dabbish, Venolia, and Cadiz}{Dabbish
  et~al\mbox{.}}{2003}]%
        {10.1145/765891.766073}
\bibfield{author}{\bibinfo{person}{Laura Dabbish}, \bibinfo{person}{Gina
  Venolia}, {and} \bibinfo{person}{JJ Cadiz}.} \bibinfo{year}{2003}\natexlab{}.
\newblock \showarticletitle{Marked for Deletion: An Analysis of Email Data}. In
  \bibinfo{booktitle}{\emph{CHI '03 Extended Abstracts on Human Factors in
  Computing Systems}} (Ft. Lauderdale, Florida, USA)
  \emph{(\bibinfo{series}{CHI EA '03})}. \bibinfo{publisher}{Association for
  Computing Machinery}, \bibinfo{address}{New York, NY, USA},
  \bibinfo{pages}{924–925}.
\newblock
\showISBNx{1581136374}
\urldef\tempurl%
\url{https://doi.org/10.1145/765891.766073}
\showDOI{\tempurl}


\bibitem[\protect\citeauthoryear{Dabbish and Kraut}{Dabbish and Kraut}{2006}]%
        {10.1145/1180875.1180941}
\bibfield{author}{\bibinfo{person}{Laura~A. Dabbish} {and}
  \bibinfo{person}{Robert~E. Kraut}.} \bibinfo{year}{2006}\natexlab{}.
\newblock \showarticletitle{Email Overload at Work: An Analysis of Factors
  Associated with Email Strain}. In \bibinfo{booktitle}{\emph{Proceedings of
  the 2006 20th Anniversary Conference on Computer Supported Cooperative Work}}
  (Banff, Alberta, Canada) \emph{(\bibinfo{series}{CSCW '06})}.
  \bibinfo{publisher}{Association for Computing Machinery},
  \bibinfo{address}{New York, NY, USA}, \bibinfo{pages}{431–440}.
\newblock
\showISBNx{1595932496}
\urldef\tempurl%
\url{https://doi.org/10.1145/1180875.1180941}
\showDOI{\tempurl}


\bibitem[\protect\citeauthoryear{Dawkins}{Dawkins}{2018}]%
        {doi:10.1080/02680513.2018.1556090}
\bibfield{author}{\bibinfo{person}{Roger Dawkins}.}
  \bibinfo{year}{2018}\natexlab{}.
\newblock \showarticletitle{Mass email at university: current literature and
  tactics for future use}.
\newblock \bibinfo{journal}{\emph{Open Learning: The Journal of Open, Distance
  and e-Learning}} \bibinfo{volume}{0}, \bibinfo{number}{0}
  (\bibinfo{year}{2018}), \bibinfo{pages}{1--17}.
\newblock
\urldef\tempurl%
\url{https://doi.org/10.1080/02680513.2018.1556090}
\showDOI{\tempurl}
\showeprint{https://doi.org/10.1080/02680513.2018.1556090}


\bibitem[\protect\citeauthoryear{Dewatripont and Tirole}{Dewatripont and
  Tirole}{2005}]%
        {dewatripont2005modes}
\bibfield{author}{\bibinfo{person}{Mathias Dewatripont} {and}
  \bibinfo{person}{Jean Tirole}.} \bibinfo{year}{2005}\natexlab{}.
\newblock \showarticletitle{Modes of communication}.
\newblock \bibinfo{journal}{\emph{Journal of political economy}}
  \bibinfo{volume}{113}, \bibinfo{number}{6} (\bibinfo{year}{2005}),
  \bibinfo{pages}{1217--1238}.
\newblock


\bibitem[\protect\citeauthoryear{Dill and Sporn}{Dill and Sporn}{1995}]%
        {dill1995emerging}
\bibfield{author}{\bibinfo{person}{David~D Dill} {and} \bibinfo{person}{Barbara
  Sporn}.} \bibinfo{year}{1995}\natexlab{}.
\newblock \bibinfo{booktitle}{\emph{Emerging Patterns of Social Demand and
  University Reform: Through a Glass Darkly. Issues in Higher Education.}}
\newblock \bibinfo{publisher}{ERIC}.
\newblock


\bibitem[\protect\citeauthoryear{Downs and Adrian}{Downs and Adrian}{2012}]%
        {downs2012assessing}
\bibfield{author}{\bibinfo{person}{Cal~W Downs} {and}
  \bibinfo{person}{Allyson~D Adrian}.} \bibinfo{year}{2012}\natexlab{}.
\newblock \bibinfo{booktitle}{\emph{Assessing organizational communication:
  Strategic communication audits}}.
\newblock \bibinfo{publisher}{Guilford Press}.
\newblock


\bibitem[\protect\citeauthoryear{Fisher, Brush, Gleave, and Smith}{Fisher
  et~al\mbox{.}}{2006}]%
        {fisher2006revisiting}
\bibfield{author}{\bibinfo{person}{Danyel Fisher}, \bibinfo{person}{AJ Brush},
  \bibinfo{person}{Eric Gleave}, {and} \bibinfo{person}{Marc~A Smith}.}
  \bibinfo{year}{2006}\natexlab{}.
\newblock \showarticletitle{Revisiting Whittaker \& Sidner's" email overload"
  ten years later}. In \bibinfo{booktitle}{\emph{Proceedings of the 2006 20th
  anniversary conference on Computer supported cooperative work}}.
  \bibinfo{pages}{309--312}.
\newblock


\bibitem[\protect\citeauthoryear{Goldhaber}{Goldhaber}{1990}]%
        {goldhaber1990organizational}
\bibfield{author}{\bibinfo{person}{Gerald~M Goldhaber}.}
  \bibinfo{year}{1990}\natexlab{}.
\newblock \bibinfo{booktitle}{\emph{Organizational communication}}.
\newblock \bibinfo{type}{{T}echnical {R}eport}. \bibinfo{institution}{WCB,}.
\newblock


\bibitem[\protect\citeauthoryear{Gray and Haahr}{Gray and Haahr}{2004}]%
        {gray2004personalised}
\bibfield{author}{\bibinfo{person}{Alan Gray} {and} \bibinfo{person}{Mads
  Haahr}.} \bibinfo{year}{2004}\natexlab{}.
\newblock \showarticletitle{Personalised, Collaborative Spam Filtering.}. In
  \bibinfo{booktitle}{\emph{CEAS}}.
\newblock


\bibitem[\protect\citeauthoryear{Greenbaum}{Greenbaum}{1974}]%
        {greenbaum1974audit}
\bibfield{author}{\bibinfo{person}{Howard~H Greenbaum}.}
  \bibinfo{year}{1974}\natexlab{}.
\newblock \showarticletitle{The audit of organizational communication}.
\newblock \bibinfo{journal}{\emph{Academy of Management Journal}}
  \bibinfo{volume}{17}, \bibinfo{number}{4} (\bibinfo{year}{1974}),
  \bibinfo{pages}{739--754}.
\newblock


\bibitem[\protect\citeauthoryear{Grevet, Choi, Kumar, and Gilbert}{Grevet
  et~al\mbox{.}}{2014}]%
        {grevet2014overload}
\bibfield{author}{\bibinfo{person}{Catherine Grevet}, \bibinfo{person}{David
  Choi}, \bibinfo{person}{Debra Kumar}, {and} \bibinfo{person}{Eric Gilbert}.}
  \bibinfo{year}{2014}\natexlab{}.
\newblock \showarticletitle{Overload is overloaded: email in the age of Gmail}.
  In \bibinfo{booktitle}{\emph{Proceedings of the sigchi conference on human
  factors in computing systems}}. \bibinfo{pages}{793--802}.
\newblock


\bibitem[\protect\citeauthoryear{Handy}{Handy}{2007}]%
        {handy2007understanding}
\bibfield{author}{\bibinfo{person}{Charles Handy}.}
  \bibinfo{year}{2007}\natexlab{}.
\newblock \bibinfo{booktitle}{\emph{Understanding organizations}}.
\newblock \bibinfo{publisher}{Penguin Uk}.
\newblock


\bibitem[\protect\citeauthoryear{Huang, Lin, and Lin}{Huang
  et~al\mbox{.}}{2009}]%
        {huang2009factors}
\bibfield{author}{\bibinfo{person}{Chien-Chih Huang},
  \bibinfo{person}{Tung-Ching Lin}, {and} \bibinfo{person}{Kuei-Ju Lin}.}
  \bibinfo{year}{2009}\natexlab{}.
\newblock \showarticletitle{Factors affecting pass-along email intentions
  (PAEIs): Integrating the social capital and social cognition theories}.
\newblock \bibinfo{journal}{\emph{Electronic Commerce Research and
  Applications}} \bibinfo{volume}{8}, \bibinfo{number}{3}
  (\bibinfo{year}{2009}), \bibinfo{pages}{160--169}.
\newblock


\bibitem[\protect\citeauthoryear{Jackson, Burgess, and Edwards}{Jackson
  et~al\mbox{.}}{2006}]%
        {jackson2006simple}
\bibfield{author}{\bibinfo{person}{Thomas~W Jackson}, \bibinfo{person}{Anthony
  Burgess}, {and} \bibinfo{person}{Janet Edwards}.}
  \bibinfo{year}{2006}\natexlab{}.
\newblock \showarticletitle{A simple approach to improving email
  communication}.
\newblock \bibinfo{journal}{\emph{Commun. ACM}} \bibinfo{volume}{49},
  \bibinfo{number}{6} (\bibinfo{year}{2006}), \bibinfo{pages}{107--109}.
\newblock


\bibitem[\protect\citeauthoryear{Jackson, Dawson, and Wilson}{Jackson
  et~al\mbox{.}}{2003}]%
        {Jackson:2003:UEI:859670.859673}
\bibfield{author}{\bibinfo{person}{Thomas~W. Jackson}, \bibinfo{person}{Ray
  Dawson}, {and} \bibinfo{person}{Darren Wilson}.}
  \bibinfo{year}{2003}\natexlab{}.
\newblock \showarticletitle{Understanding Email Interaction Increases
  Organizational Productivity}.
\newblock \bibinfo{journal}{\emph{Commun. ACM}} \bibinfo{volume}{46},
  \bibinfo{number}{8} (\bibinfo{date}{Aug.} \bibinfo{year}{2003}),
  \bibinfo{pages}{80--84}.
\newblock
\showISSN{0001-0782}
\urldef\tempurl%
\url{https://doi.org/10.1145/859670.859673}
\showDOI{\tempurl}


\bibitem[\protect\citeauthoryear{Katz and Kahn}{Katz and Kahn}{2008}]%
        {katz2008communication}
\bibfield{author}{\bibinfo{person}{Daniel Katz} {and} \bibinfo{person}{Robert~L
  Kahn}.} \bibinfo{year}{2008}\natexlab{}.
\newblock \showarticletitle{Communication: the flow of information}.
\newblock \bibinfo{journal}{\emph{Communication theory}}
  (\bibinfo{year}{2008}), \bibinfo{pages}{382--389}.
\newblock


\bibitem[\protect\citeauthoryear{Khan}{Khan}{2017}]%
        {khan2017social}
\bibfield{author}{\bibinfo{person}{M~Laeeq Khan}.}
  \bibinfo{year}{2017}\natexlab{}.
\newblock \showarticletitle{Social media engagement: What motivates user
  participation and consumption on YouTube?}
\newblock \bibinfo{journal}{\emph{Computers in human behavior}}
  \bibinfo{volume}{66} (\bibinfo{year}{2017}), \bibinfo{pages}{236--247}.
\newblock


\bibitem[\protect\citeauthoryear{Klimt and Yang}{Klimt and Yang}{2004}]%
        {klimt2004enron}
\bibfield{author}{\bibinfo{person}{Bryan Klimt} {and} \bibinfo{person}{Yiming
  Yang}.} \bibinfo{year}{2004}\natexlab{}.
\newblock \showarticletitle{The enron corpus: A new dataset for email
  classification research}. In \bibinfo{booktitle}{\emph{European Conference on
  Machine Learning}}. Springer, \bibinfo{pages}{217--226}.
\newblock


\bibitem[\protect\citeauthoryear{Kueng}{Kueng}{2000}]%
        {kueng2000process}
\bibfield{author}{\bibinfo{person}{Peter Kueng}.}
  \bibinfo{year}{2000}\natexlab{}.
\newblock \showarticletitle{Process performance measurement system: a tool to
  support process-based organizations}.
\newblock \bibinfo{journal}{\emph{Total quality management}}
  \bibinfo{volume}{11}, \bibinfo{number}{1} (\bibinfo{year}{2000}),
  \bibinfo{pages}{67--85}.
\newblock


\bibitem[\protect\citeauthoryear{Kuh and Whitt}{Kuh and Whitt}{2000}]%
        {kuh2000culture}
\bibfield{author}{\bibinfo{person}{GD Kuh} {and} \bibinfo{person}{EJ Whitt}.}
  \bibinfo{year}{2000}\natexlab{}.
\newblock \bibinfo{title}{Culture in American colleges and universities.
  Organizational \& Governance in Higher Education. MC Brown, Ed}.
\newblock
\newblock


\bibitem[\protect\citeauthoryear{Lazar, Feng, and Hochheiser}{Lazar
  et~al\mbox{.}}{2017}]%
        {lazar2017research}
\bibfield{author}{\bibinfo{person}{Jonathan Lazar},
  \bibinfo{person}{Jinjuan~Heidi Feng}, {and} \bibinfo{person}{Harry
  Hochheiser}.} \bibinfo{year}{2017}\natexlab{}.
\newblock \bibinfo{booktitle}{\emph{Research methods in human-computer
  interaction}}.
\newblock \bibinfo{publisher}{Morgan Kaufmann}.
\newblock


\bibitem[\protect\citeauthoryear{Lu, Wen, Pan, and Lai}{Lu
  et~al\mbox{.}}{2012}]%
        {lu2012epic}
\bibfield{author}{\bibinfo{person}{Jie Lu}, \bibinfo{person}{Zhen Wen},
  \bibinfo{person}{Shimei Pan}, {and} \bibinfo{person}{Jennifer Lai}.}
  \bibinfo{year}{2012}\natexlab{}.
\newblock \showarticletitle{EPIC: a multi-tiered approach to enterprise email
  prioritization}. In \bibinfo{booktitle}{\emph{Proceedings of the 2012 ACM
  international conference on Intelligent User Interfaces}}.
  \bibinfo{pages}{199--202}.
\newblock


\bibitem[\protect\citeauthoryear{M{\"a}ki, J{\"a}rvenp{\"a}{\"a}, Ziegler, and
  Ziegler}{M{\"a}ki et~al\mbox{.}}{2004}]%
        {maki2004communication}
\bibfield{author}{\bibinfo{person}{Eerikki M{\"a}ki}, \bibinfo{person}{Eila
  J{\"a}rvenp{\"a}{\"a}}, \bibinfo{person}{Kirsi Ziegler}, {and}
  \bibinfo{person}{Kirsi Ziegler}.} \bibinfo{year}{2004}\natexlab{}.
\newblock \showarticletitle{Communication and knowledge sharing in a
  decentralized organization}.
\newblock  (\bibinfo{year}{2004}).
\newblock


\bibitem[\protect\citeauthoryear{Malone}{Malone}{1987}]%
        {malone1987information}
\bibfield{author}{\bibinfo{person}{T Malone}.} \bibinfo{year}{1987}\natexlab{}.
\newblock \showarticletitle{The information lens: An intelligent system for
  sharing information in organizations}.
\newblock \bibinfo{journal}{\emph{Proceedings, Computer Human Interaction,
  Human Factors in Computing Systems}} (\bibinfo{year}{1987}).
\newblock


\bibitem[\protect\citeauthoryear{Mark, Iqbal, Czerwinski, Johns, Sano, and
  Lutchyn}{Mark et~al\mbox{.}}{2016}]%
        {mark2016email}
\bibfield{author}{\bibinfo{person}{Gloria Mark}, \bibinfo{person}{Shamsi~T
  Iqbal}, \bibinfo{person}{Mary Czerwinski}, \bibinfo{person}{Paul Johns},
  \bibinfo{person}{Akane Sano}, {and} \bibinfo{person}{Yuliya Lutchyn}.}
  \bibinfo{year}{2016}\natexlab{}.
\newblock \showarticletitle{Email duration, batching and self-interruption:
  Patterns of email use on productivity and stress}. In
  \bibinfo{booktitle}{\emph{Proceedings of the 2016 CHI Conference on Human
  Factors in Computing Systems}}. \bibinfo{pages}{1717--1728}.
\newblock


\bibitem[\protect\citeauthoryear{Masland}{Masland}{1985}]%
        {masland1985organizational}
\bibfield{author}{\bibinfo{person}{Andrew~T Masland}.}
  \bibinfo{year}{1985}\natexlab{}.
\newblock \showarticletitle{Organizational culture in the study of higher
  education}.
\newblock \bibinfo{journal}{\emph{The Review of Higher Education}}
  \bibinfo{volume}{8}, \bibinfo{number}{2} (\bibinfo{year}{1985}),
  \bibinfo{pages}{157--168}.
\newblock


\bibitem[\protect\citeauthoryear{Merten and Gloor}{Merten and Gloor}{2010}]%
        {merten2010too}
\bibfield{author}{\bibinfo{person}{Frank Merten} {and} \bibinfo{person}{Peter
  Gloor}.} \bibinfo{year}{2010}\natexlab{}.
\newblock \showarticletitle{Too much e-mail decreases job satisfaction}.
\newblock \bibinfo{journal}{\emph{Procedia-Social and Behavioral Sciences}}
  \bibinfo{volume}{2}, \bibinfo{number}{4} (\bibinfo{year}{2010}),
  \bibinfo{pages}{6457--6465}.
\newblock


\bibitem[\protect\citeauthoryear{Mishra, Boynton, and Mishra}{Mishra
  et~al\mbox{.}}{2014}]%
        {mishra2014driving}
\bibfield{author}{\bibinfo{person}{Karen Mishra}, \bibinfo{person}{Lois
  Boynton}, {and} \bibinfo{person}{Aneil Mishra}.}
  \bibinfo{year}{2014}\natexlab{}.
\newblock \showarticletitle{Driving employee engagement: The expanded role of
  internal communications}.
\newblock \bibinfo{journal}{\emph{International Journal of Business
  Communication}} \bibinfo{volume}{51}, \bibinfo{number}{2}
  (\bibinfo{year}{2014}), \bibinfo{pages}{183--202}.
\newblock


\bibitem[\protect\citeauthoryear{Myers and Myers}{Myers and Myers}{1982}]%
        {myers1982managing}
\bibfield{author}{\bibinfo{person}{Michele~Tolela Myers} {and}
  \bibinfo{person}{Gail~E Myers}.} \bibinfo{year}{1982}\natexlab{}.
\newblock \bibinfo{booktitle}{\emph{Managing by communication: An
  organizational approach}}.
\newblock \bibinfo{publisher}{McGraw-Hill College}.
\newblock


\bibitem[\protect\citeauthoryear{Norris, White, Nowell, Mrklas, and
  Stelfox}{Norris et~al\mbox{.}}{2017}]%
        {norris2017stakeholders}
\bibfield{author}{\bibinfo{person}{Jill~M Norris}, \bibinfo{person}{Deborah~E
  White}, \bibinfo{person}{Lorelli Nowell}, \bibinfo{person}{Kelly Mrklas},
  {and} \bibinfo{person}{Henry~T Stelfox}.} \bibinfo{year}{2017}\natexlab{}.
\newblock \showarticletitle{How do stakeholders from multiple hierarchical
  levels of a large provincial health system define engagement? A qualitative
  study}.
\newblock \bibinfo{journal}{\emph{Implementation Science}}
  \bibinfo{volume}{12}, \bibinfo{number}{1} (\bibinfo{year}{2017}),
  \bibinfo{pages}{98}.
\newblock


\bibitem[\protect\citeauthoryear{Odine}{Odine}{2015}]%
        {odine2015communication}
\bibfield{author}{\bibinfo{person}{Maurice Odine}.}
  \bibinfo{year}{2015}\natexlab{}.
\newblock \showarticletitle{Communication problems in management}.
\newblock \bibinfo{journal}{\emph{Journal of Emerging Issues in Economics,
  Finance, and Banking (JEIEFB)}} \bibinfo{volume}{4}, \bibinfo{number}{2}
  (\bibinfo{year}{2015}).
\newblock


\bibitem[\protect\citeauthoryear{Paczkowski and Kuruzovich}{Paczkowski and
  Kuruzovich}{2016}]%
        {paczkowski2016checking}
\bibfield{author}{\bibinfo{person}{William~F Paczkowski} {and}
  \bibinfo{person}{Jason Kuruzovich}.} \bibinfo{year}{2016}\natexlab{}.
\newblock \showarticletitle{Checking email in the bathroom: monitoring email
  responsiveness behavior in the workplace}.
\newblock \bibinfo{journal}{\emph{American Journal of Management}}
  \bibinfo{volume}{16}, \bibinfo{number}{2} (\bibinfo{year}{2016}).
\newblock


\bibitem[\protect\citeauthoryear{Rasmussen and Ulrich}{Rasmussen and
  Ulrich}{2015}]%
        {rasmussen2015learning}
\bibfield{author}{\bibinfo{person}{Thomas Rasmussen} {and}
  \bibinfo{person}{Dave Ulrich}.} \bibinfo{year}{2015}\natexlab{}.
\newblock \showarticletitle{Learning from practice: how HR analytics avoids
  being a management fad}.
\newblock \bibinfo{journal}{\emph{Organizational Dynamics}}
  \bibinfo{volume}{44}, \bibinfo{number}{3} (\bibinfo{year}{2015}),
  \bibinfo{pages}{236--242}.
\newblock


\bibitem[\protect\citeauthoryear{Raven and Flanders}{Raven and
  Flanders}{1996}]%
        {10.1145/227614.227615}
\bibfield{author}{\bibinfo{person}{Mary~Elizabeth Raven} {and}
  \bibinfo{person}{Alicia Flanders}.} \bibinfo{year}{1996}\natexlab{}.
\newblock \showarticletitle{Using Contextual Inquiry to Learn about Your
  Audiences}.
\newblock \bibinfo{journal}{\emph{SIGDOC Asterisk J. Comput. Doc.}}
  \bibinfo{volume}{20}, \bibinfo{number}{1} (\bibinfo{date}{Feb.}
  \bibinfo{year}{1996}), \bibinfo{pages}{1–13}.
\newblock
\showISSN{0731-1001}
\urldef\tempurl%
\url{https://doi.org/10.1145/227614.227615}
\showDOI{\tempurl}


\bibitem[\protect\citeauthoryear{Reeves, Roy, Gorman, and Morley}{Reeves
  et~al\mbox{.}}{2008}]%
        {reeves2008marketplace}
\bibfield{author}{\bibinfo{person}{Byron Reeves}, \bibinfo{person}{Simon Roy},
  \bibinfo{person}{Brian Gorman}, {and} \bibinfo{person}{Teresa Morley}.}
  \bibinfo{year}{2008}\natexlab{}.
\newblock \showarticletitle{A marketplace for attention: Responses to a
  synthetic currency used to signal information importance in e-mail}.
\newblock \bibinfo{journal}{\emph{First Monday}} \bibinfo{volume}{13},
  \bibinfo{number}{5} (\bibinfo{year}{2008}).
\newblock


\bibitem[\protect\citeauthoryear{Sarrafzadeh, Awadallah, Lin, Lee, Shokouhi,
  and Dumais}{Sarrafzadeh et~al\mbox{.}}{2019}]%
        {Sarrafzadeh2019CharacterizingAP}
\bibfield{author}{\bibinfo{person}{Bahareh Sarrafzadeh},
  \bibinfo{person}{Ahmed~Hassan Awadallah}, \bibinfo{person}{Christopher~H.
  Lin}, \bibinfo{person}{Chia-Jung Lee}, \bibinfo{person}{Milad Shokouhi},
  {and} \bibinfo{person}{Susan~T. Dumais}.} \bibinfo{year}{2019}\natexlab{}.
\newblock \showarticletitle{Characterizing and Predicting Email Deferral
  Behavior}. In \bibinfo{booktitle}{\emph{WSDM '19}}.
\newblock


\bibitem[\protect\citeauthoryear{Schuler}{Schuler}{1979}]%
        {SCHULER1979268}
\bibfield{author}{\bibinfo{person}{Randall~S. Schuler}.}
  \bibinfo{year}{1979}\natexlab{}.
\newblock \showarticletitle{A role perception transactional process model for
  organizational communication-outcome relationships}.
\newblock \bibinfo{journal}{\emph{Organizational Behavior and Human
  Performance}} \bibinfo{volume}{23}, \bibinfo{number}{2}
  (\bibinfo{year}{1979}), \bibinfo{pages}{268 -- 291}.
\newblock
\showISSN{0030-5073}
\urldef\tempurl%
\url{https://doi.org/10.1016/0030-5073(79)90058-8}
\showDOI{\tempurl}


\bibitem[\protect\citeauthoryear{Shriberg et~al\mbox{.}}{Shriberg
  et~al\mbox{.}}{2002}]%
        {shriberg2002sustainability}
\bibfield{author}{\bibinfo{person}{Michael~P Shriberg} {et~al\mbox{.}}}
  \bibinfo{year}{2002}\natexlab{}.
\newblock \emph{\bibinfo{title}{Sustainability in US higher education:
  organizational factors influencing campus environmental performance and
  leadership}}.
\newblock \bibinfo{thesistype}{Ph.D. Dissertation}. \bibinfo{school}{University
  of Michigan Ann Arbor, MI}.
\newblock


\bibitem[\protect\citeauthoryear{Sproull and Kiesler}{Sproull and
  Kiesler}{1991}]%
        {sproull1991computers}
\bibfield{author}{\bibinfo{person}{Lee Sproull} {and} \bibinfo{person}{Sara
  Kiesler}.} \bibinfo{year}{1991}\natexlab{}.
\newblock \showarticletitle{Computers, networks and work}.
\newblock \bibinfo{journal}{\emph{Scientific American}} \bibinfo{volume}{265},
  \bibinfo{number}{3} (\bibinfo{year}{1991}), \bibinfo{pages}{116--127}.
\newblock


\bibitem[\protect\citeauthoryear{Stohl}{Stohl}{1995}]%
        {stohl1995organizational}
\bibfield{author}{\bibinfo{person}{Cynthia Stohl}.}
  \bibinfo{year}{1995}\natexlab{}.
\newblock \bibinfo{booktitle}{\emph{Organizational communication}}.
\newblock Number~5. \bibinfo{publisher}{Sage}.
\newblock


\bibitem[\protect\citeauthoryear{Stohl and Redding}{Stohl and Redding}{1987}]%
        {stohl1987messages}
\bibfield{author}{\bibinfo{person}{Cynthia Stohl} {and}
  \bibinfo{person}{W~Charles Redding}.} \bibinfo{year}{1987}\natexlab{}.
\newblock \showarticletitle{Messages and message exchange processes.}
\newblock  (\bibinfo{year}{1987}).
\newblock


\bibitem[\protect\citeauthoryear{Swanborn}{Swanborn}{2010}]%
        {swanborn2010case}
\bibfield{author}{\bibinfo{person}{Peter Swanborn}.}
  \bibinfo{year}{2010}\natexlab{}.
\newblock \bibinfo{booktitle}{\emph{Case study research: What, why and how?}}
\newblock \bibinfo{publisher}{Sage}.
\newblock


\bibitem[\protect\citeauthoryear{Trespalacios and Perkins}{Trespalacios and
  Perkins}{2016}]%
        {trespalacios2016effects}
\bibfield{author}{\bibinfo{person}{Jes{\'u}s~H Trespalacios} {and}
  \bibinfo{person}{Ross~A Perkins}.} \bibinfo{year}{2016}\natexlab{}.
\newblock \showarticletitle{Effects of personalization and invitation email
  length on web-based survey response rates}.
\newblock \bibinfo{journal}{\emph{TechTrends}} \bibinfo{volume}{60},
  \bibinfo{number}{4} (\bibinfo{year}{2016}), \bibinfo{pages}{330--335}.
\newblock


\bibitem[\protect\citeauthoryear{Van~Vleck}{Van~Vleck}{2012}]%
        {van2012electronic}
\bibfield{author}{\bibinfo{person}{Tom Van~Vleck}.}
  \bibinfo{year}{2012}\natexlab{}.
\newblock \showarticletitle{Electronic mail and text messaging in CTSS,
  1965-1973}.
\newblock \bibinfo{journal}{\emph{IEEE Annals of the History of Computing}}
  \bibinfo{volume}{34}, \bibinfo{number}{1} (\bibinfo{year}{2012}),
  \bibinfo{pages}{4--6}.
\newblock


\bibitem[\protect\citeauthoryear{Wainer, Dabbish, and Kraut}{Wainer
  et~al\mbox{.}}{2011}]%
        {10.1145/1978942.1979456}
\bibfield{author}{\bibinfo{person}{Jaclyn Wainer}, \bibinfo{person}{Laura
  Dabbish}, {and} \bibinfo{person}{Robert Kraut}.}
  \bibinfo{year}{2011}\natexlab{}.
\newblock \showarticletitle{Should I Open This Email? Inbox-Level Cues,
  Curiosity and Attention to Email}. In \bibinfo{booktitle}{\emph{Proceedings
  of the SIGCHI Conference on Human Factors in Computing Systems}} (Vancouver,
  BC, Canada) \emph{(\bibinfo{series}{CHI '11})}.
  \bibinfo{publisher}{Association for Computing Machinery},
  \bibinfo{address}{New York, NY, USA}, \bibinfo{pages}{3439–3448}.
\newblock
\showISBNx{9781450302289}
\urldef\tempurl%
\url{https://doi.org/10.1145/1978942.1979456}
\showDOI{\tempurl}


\bibitem[\protect\citeauthoryear{Whittaker and Sidner}{Whittaker and
  Sidner}{1996}]%
        {whittaker1996email}
\bibfield{author}{\bibinfo{person}{Steve Whittaker} {and}
  \bibinfo{person}{Candace Sidner}.} \bibinfo{year}{1996}\natexlab{}.
\newblock \showarticletitle{Email overload: exploring personal information
  management of email}. In \bibinfo{booktitle}{\emph{Proceedings of the SIGCHI
  conference on Human factors in computing systems}}.
  \bibinfo{pages}{276--283}.
\newblock


\bibitem[\protect\citeauthoryear{Yang, Dumais, Bennett, and Awadallah}{Yang
  et~al\mbox{.}}{2017}]%
        {yang2017characterizing}
\bibfield{author}{\bibinfo{person}{Liu Yang}, \bibinfo{person}{Susan~T Dumais},
  \bibinfo{person}{Paul~N Bennett}, {and} \bibinfo{person}{Ahmed~Hassan
  Awadallah}.} \bibinfo{year}{2017}\natexlab{}.
\newblock \showarticletitle{Characterizing and predicting enterprise email
  reply behavior}. In \bibinfo{booktitle}{\emph{Proceedings of the 40th
  International ACM SIGIR Conference on Research and Development in Information
  Retrieval}}. \bibinfo{pages}{235--244}.
\newblock


\bibitem[\protect\citeauthoryear{Yin}{Yin}{2003}]%
        {yin2003case}
\bibfield{author}{\bibinfo{person}{Robert~K Yin}.}
  \bibinfo{year}{2003}\natexlab{}.
\newblock \showarticletitle{Case study research: Design and methods (Vol. 5)}.
\newblock  (\bibinfo{year}{2003}).
\newblock


\end{thebibliography}

\section{Appendices}

\appendix

\section{Communicator Interview Protocol}

See Table \ref{tab:gatekeeper_protocol}.
\begin{table}[!htbp]
\small
\centering
\arrayrulecolor[rgb]{0.8,0.8,0.8}
\resizebox{\textwidth}{!}{%
\begin{tabular}{|l|r|p{13cm}|} 
\hline
\textbf{Part}  & \multicolumn{1}{l|}{} & \textbf{Questions}                                                                                                                                                  \\ 
\arrayrulecolor{black}\hline
1.Practice     & 1                     & How do you send bulk emails for your unit and why do you send it in that way      ?                                                          \\ 
\arrayrulecolor[rgb]{0.8,0.8,0.8}\hline
               & \multicolumn{1}{l|}{} & Who are your client? What are their requests? Do they send emails themselves?                                                                                          \\ 
\hline
               & \multicolumn{1}{l|}{} & Who offer the mailing list and content?                                                                                                                             \\ 
\hline
               & 2                     & When do you send bulk emails to all employees versus subgroups?                                                                                                                 \\ 
\hline
               & 3                     & When do you decide to put a message in a bulletin of newsletter versus an individual bulk email?                                                                            \\ 
\hline
               & 4                     & Besides newsletters and individual bulk emails, what are the other mechanism you use to communicate? \\ 
\hline
               & 5                     & What happened with the bulk emails you sent? Are you aware of how often they read or do they understand it carefully?                                                  \\ 
\hline
2. Email Cases & 1                     & The title, goal, and recipient of the email                                                                                                                          \\ 
\hline
               & 2                     & Which channel did you use to send it?                                                                                     \\ 
\hline
               & 3                     & Who do you imagine should read this? Who do you imagine did read this? Are they the same?                                                                           \\ 
\hline
               & 4                     & Is who should read this the same as who the email was sent to?                                                                                                 \\ 
\hline
               & 5                     & Who should read in detail and who can just scan it?                                                                                                                 \\ 
\hline
               & 6                     & Who do you imagine should take actions (click links, take surveys, reply)? Did they take the actions?                                                                \\ 
\hline
               & 7                     & Do you have experience that some recipients asked about/forgot messages that you've already sent in an email? When will you send it for multiple times?                  \\ 

\hline
3. Assessment   & 1                     & What's your sense of how email communication between university and employees work (well or poorly)?                                                                \\
\hline
\end{tabular}}
\arrayrulecolor{black}
\caption{Interview Protocol of Communicators.}
\label{tab:gatekeeper_protocol}
\end{table}

\section{Recipients Interview Protocol}
See Table \ref{tab:command} for search commands for different types of emails. The following are the questions we asked the recipient for each email:
\begin{itemize}
    \item Did you label/trash this email? Did you open this email or leave it unread? Why?
    \item Did you recognize the sender?
    \item If you opened it, did you scan or read it in detail? Why?
    \item If you didn't open it, open it now, do you find anything important to you?
    \item Did this email require actions? Did you take actions and how soon did you take? Why?
    \item Rate the importance/urgency/relevance of this email from 1 (lowest) - 5 (highest).
\end{itemize}

\begin{table}[!htbp]
\centering
\footnotesize
\begin{tabular}{|p{5cm}|p{8cm}|} 
\hline
 \textbf{Questions}                                                                                  & \textbf{Command}                                                                                                                                                                                                          \\ 
\hline
How many emails did the participant receive within 1 week?                                      & newer\_than:7d,in:anywhere AND NOT from:me                                                                                                                                                                                 \\ 
\hline
How many emails did the participant receive and not read within 1 week?                              & newer\_than:7d,in:anywhere AND label:unread AND NOT from:me                                                                                                                                                                \\ 
\hline
How many emails did the participant receive from their organizations within 1 week?                  & newer\_than:7d from:umn.edu AND NOT from:me ,in:anywhere                                                                                                                                                                   \\ 
\hline
How many emails did the participant receive from their organizations and unread within 1 week?       & newer\_than:7d,in:anywhere AND NOT from:me AND label:unread AND from:umn.edu                                                                                                                                               \\ 
\hline
How many mass emails did the participant receive from the organization within 1 week?                & newer\_than:7d,from:umn.edu AND NOT from:me AND (list:(local) OR list:(list) OR to:(lists) OR (category:(Forums\textbar{}Promotions) )) ,in:anywhere                                                                       \\ 
\hline
How many mass emails did the participant receive from the organization and unread within 1 week?     & newer\_than:7d,from:umn.edu AND NOT from:me AND label:unread AND (list:(local) OR list:(list) OR to:(lists) OR (category:(Forums\textbar{}Promotions))) ,in:anywhere                                                       \\ 
\hline
How many newsletters did the participant receive from the organization within 1 week?                & newer\_than:7d,from:umn.edu AND NOT from:me AND (list:(local) OR list:(list) OR to:(lists) OR (category:(Forums\textbar{}Promotions) )) AND (subject:(news\textbar{}update\textbar{}brief)) ,in:anywhere                   \\ 
\hline
How many newsletters did the participant receive from the organization and unread within 1 week?     & newer\_than:7d,from:umn.edu AND NOT from:me AND label:unread AND (list:(local) OR list:(list) OR to:(lists) OR (category:(Forums\textbar{}Promotions))) AND (subject:(news\textbar{}update\textbar{}brief)) ,in:anywhere  \\ 
\hline
How many personal emails did the participant receive from the organization within 1 week?            & newer\_than:7d,from:umn.edu AND NOT from:me AND to:umn.edu AND NOT (list:(local) OR list:(list) OR to:lists OR (category:(Forums\textbar{}Promotions))) ,in:anywhere                                                      \\ 
\hline
How many personal emails did the participant receive from the organization and unread within 1 week? & newer\_than:7d,from:umn.edu AND NOT from:me AND to:umn.edu AND NOT (list:(local) OR list:(list) OR to:lists OR (category:(Forums\textbar{}Promotions))) ,in:anywhere ,label:unread                                        \\
\hline
\end{tabular}
\caption{Email searching commands for Gmail. The University of Minnesota used a Gmail system based on G Suite for Education.}
\label{tab:command}
\end{table}
\end{document}